\documentclass{elsart}
\journal{Physics Reports}
\usepackage{graphicx}
\usepackage{amssymb}

\begin{document}


\newpage

\begin{frontmatter}
\title{Melting and nonmelting of solid surfaces and nanosystems}
\author[sissa,democritos]{U. Tartaglino},
\author[sissa,democritos]{T. Zykova-Timan\corauthref{corresponding}}
\ead{tzykova@sissa.it}
\corauth[corresponding]{Corresponding author.}
\author[udine,democritos]{F. Ercolessi}
\author[sissa,democritos,ictp]{E. Tosatti}
\address[sissa]{ International School for Advanced Studies (SISSA-ISAS),
  via Beirut 2, 34014 Trieste, Italy}
\address[democritos]{INFM Democritos National Simulation Center, 
  Trieste, Italy}
\address[udine]{ Dipartimento di Fisica, Universit\`a di Udine,
  Via delle Scienze, 208, I-33100 Udine, Italy}
\address[ictp]{International Center for Theoretical Physics (ICTP),
  Strada Costiera 11, 34014, Trieste, Italy}


\begin{abstract}
We present  an extensive but concise review of our present understanding,
largely based on theory and simulation work from our group, on
the equilibrium behavior of solid surfaces and nanosystems close to the
bulk melting point. In the first part we define phenomena,
in particular surface melting and nonmelting, and review
some related theoretical approaches, from heuristic theories
to computer simulation. In the second part we describe
the surface melting/nonmelting behavior
of several different classes of
solids, ranging from van der Waals crystals, to valence
semiconductors, to ionic crystals and metals.
In the third part, we address special cases such as strained solids,
the defreezing of glass surfaces, and rotational surface melting.
Next, we digress briefly to surface layering
of a liquid metal, possibly leading to solid-like or hexatic
two dimensional phases floating on the liquid. In the final part, the
relationship of surface melting to the premelting of nanoclusters
and nanowires is reviewed.
\end{abstract}

\begin{keyword}
Surface melting; Surface phase transitions; Surface thermodynamics;
Wetting; Molecular dynamics simulation.

\emph{Pacs:}\ \  61.30Hn; 64.70Dv; 68.08.-p;68.35.Md; 61.46.+w 
\end{keyword}

\end{frontmatter}
\newpage
\tableofcontents

\clearpage
\listoffigures

\newpage 
\section{Introduction: surface melting/nonmelting}

Melting, certainly one of the longest known phase transitions,
always appeared a very special one. Although undoubtedly first order,
it curiously displays only {\em half} the hysteresis cycle one would 
expect. In a regular first order phase transition, the free energies 
of two phases cross, while both remain locally stable beyond the 
crossing. As a result, it is generally possible to undercool the high 
temperature  phase, and to overheat the low temperature phase.
Periodic heating and cooling through a first order transition
will thus generate a full hysteresis cycle.

In melting, this is only half true. It is generally possible
and easy to supercool a liquid: a glass of water may not freeze at all
during an icy night, only to do that suddenly when disturbed the next
morning.  On the contrary, it is surprisingly difficult and often
impossible to overheat a free solid. The reason for that is that a free
solid has surfaces; and when the melting temperature is approached,
melting of the bulk crystal is ready to begin from there. Supercooling 
of the liquid is allowed by the absence of a solid germ; overheating of
the solid is prevented by that omnipresent germ, a wet surface. This
observation provides perhaps the first macroscopic hint that solid surfaces
might be wet already somewhat below the melting point. The starting
motivation for a microscopic study of surface melting begins right
there.
  
\begin{figure}[bh]
  \includegraphics[width=\textwidth]{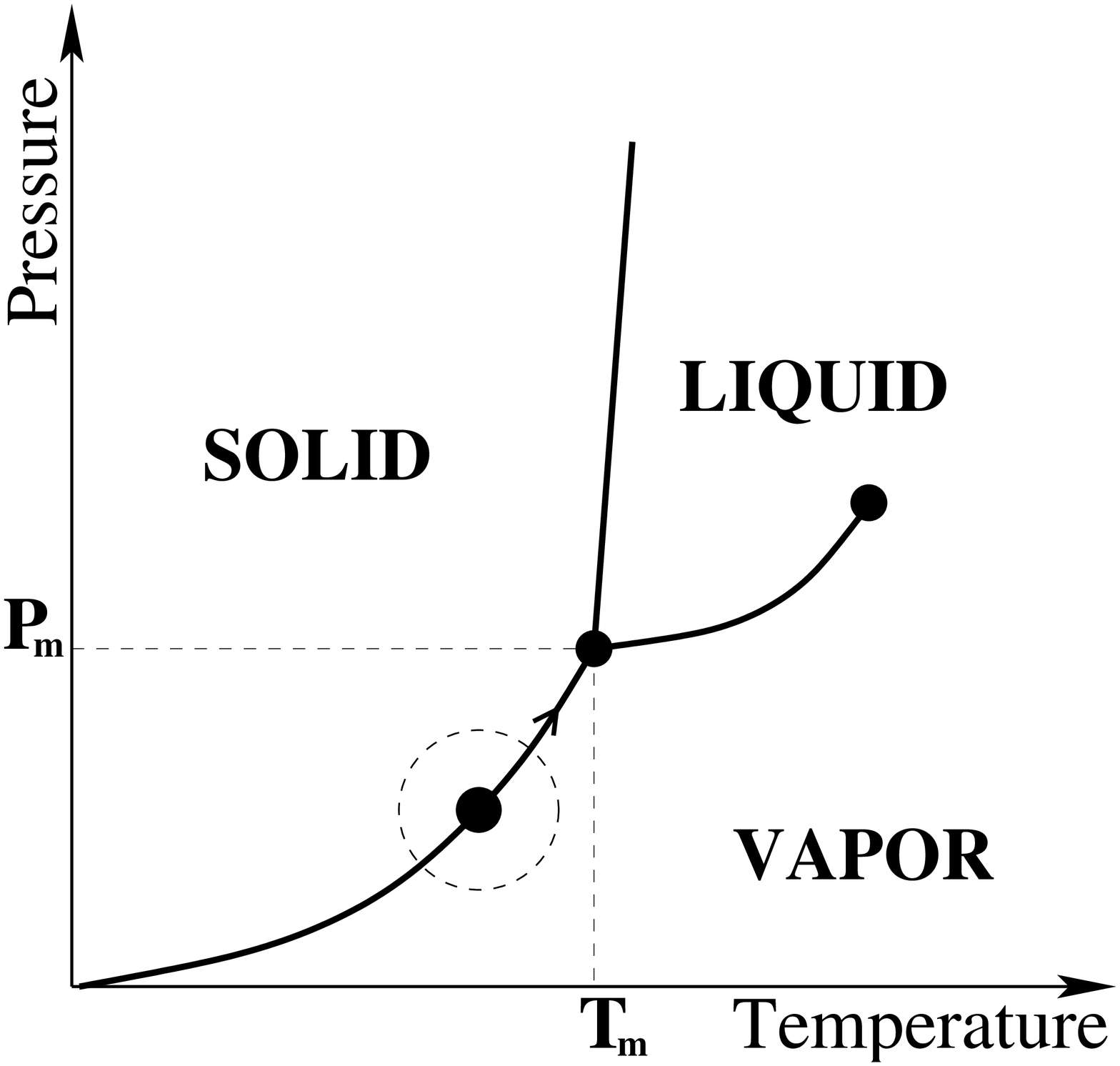}
 \caption[Melting of a pure and homogeneous material\ldots]{
           \label{pt-plane-fig}
           Melting of a pure and homogeneous material by heating along the
           solid-vapor coexistence curve up to the triple point.}
\end{figure}

\begin{figure}
  \includegraphics[width=\textwidth]{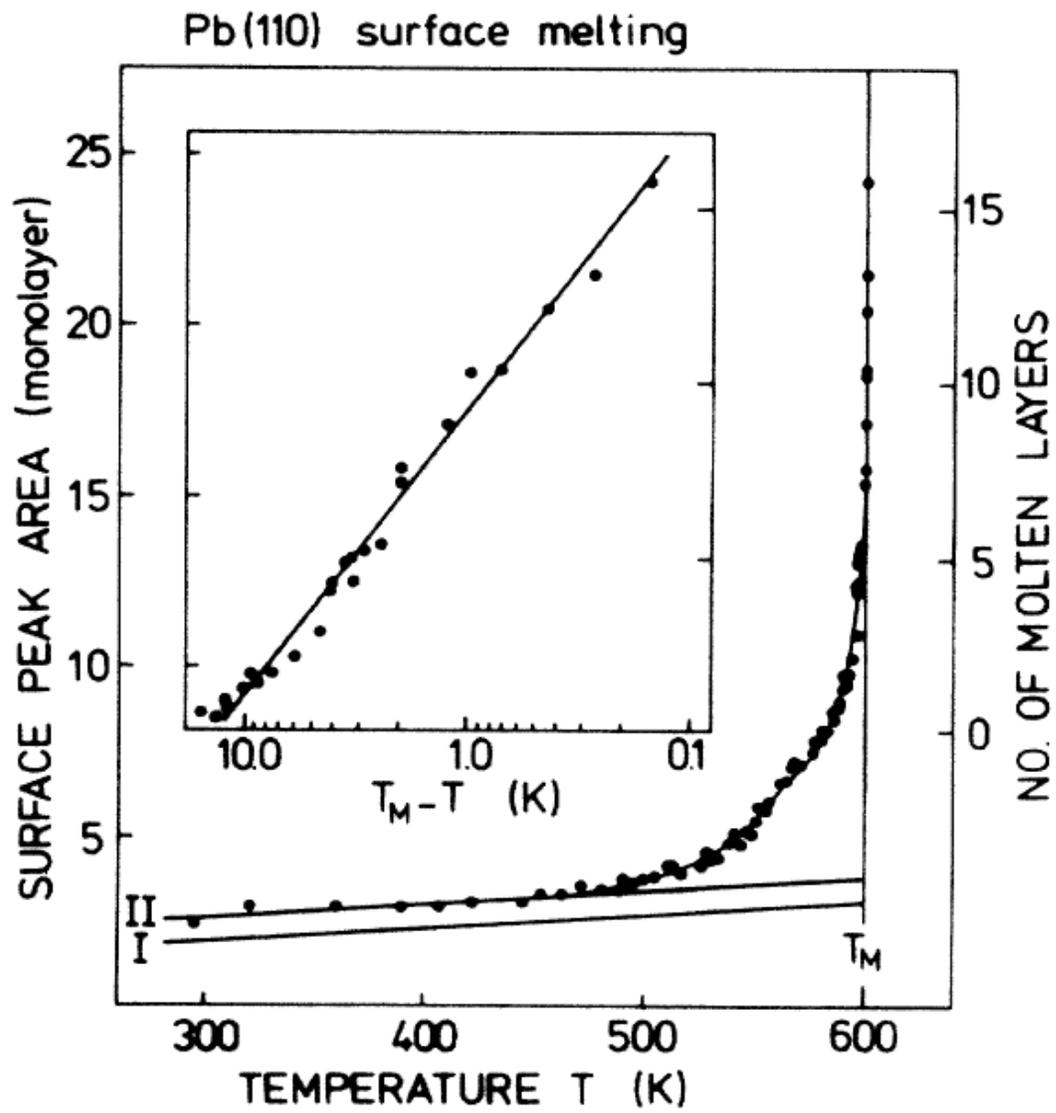}
    \caption[Surface melting of Pb(110)\ldots]{
           \label{pb110-meis-fig}
           Experimental evidence of surface melting of Pb(110): Medium energy
           ion scattering shows an increasing number of disordered (liquid)
           layers approaching the melting temperature. From
           Ref.\,\protect\cite{frenken86}.}
\end{figure}

Consider a semi-infinite homogeneous solid, e.g., a crystal with a 
free surface, in thermodynamic equilibrium at temperature $T$ and 
pressure $P$ with its own vapor (Fig.\,\ref{pt-plane-fig}).
As $T$ and $P$ are raised to approach melting at the bulk triple 
point ($T_m$, $P_m$), where solid, gas and liquid coexist, 
the solid-vapor interface will generally wet itself with an atomically 
thin liquid film, whose thickness $\ell(T) \to \infty$ when $T \to T_m$. 
This is called surface melting (SM). A rather famous microscopic 
characterization of SM is that of Pb(110), first reported by Frenken 
and van der Veen\,\cite{frenken86}. Medium energy ion scattering 
data reveal (Fig.\,\ref{pb110-meis-fig}) the presence at the solid-vapor 
interface of a disordered (liquid) film  whose thickness $\ell(T)$
grows without limit as  $T \to T_m$.
A number of case studies are now
known in the literature, and SM turns out to be the common behavior for many 
surfaces. 
It is however not general, and many surfaces are known that do not melt in the 
same sense, but remain solid and crystalline all the way up to $T_m$.
Not surprisingly, this behavior has been called surface nonmelting
(NM)\,\cite{carnevali}. Slightly more surprisingly, SM and NM can both occur for
the same substance, of course on two different crystallographic faces. For
example, NM was demonstrated for Pb(111)\,\cite{frenken87}, while by contrast 
SM prevails on Pb(110)\,\cite{frenken86}.

\begin{figure}
\includegraphics[width=\textwidth]{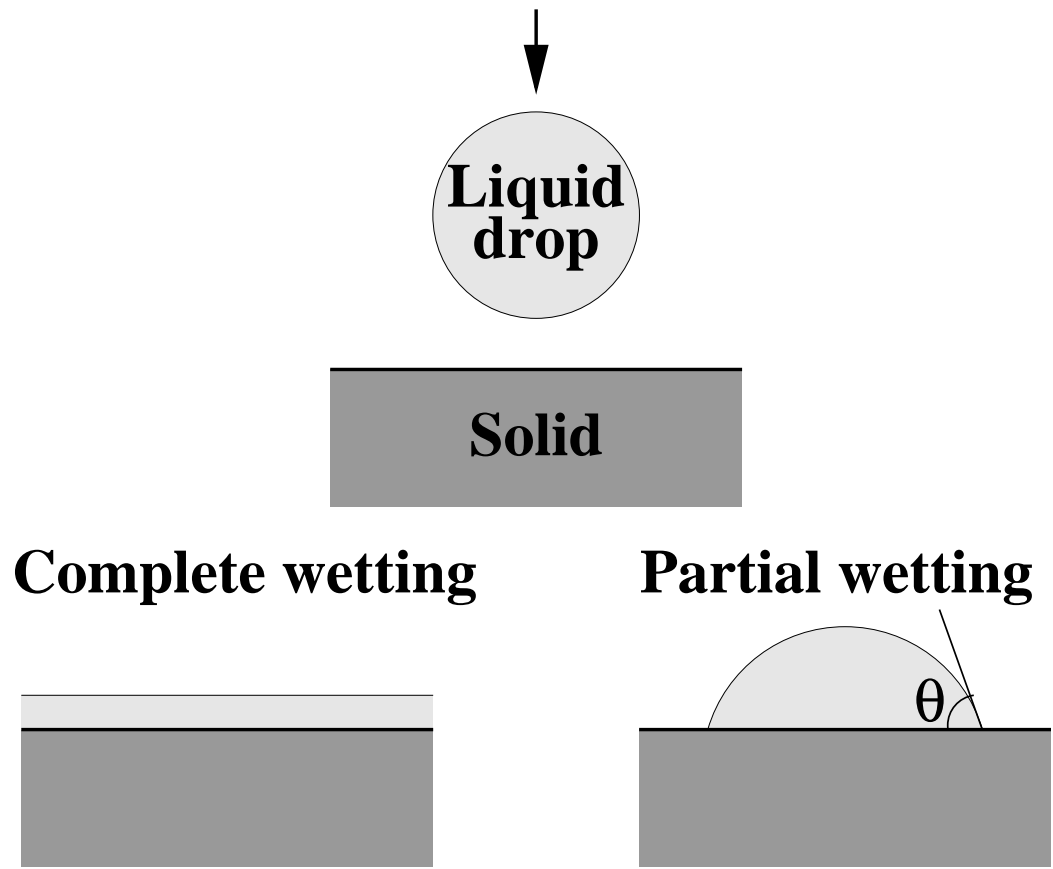}
    \caption[Partial wetting\ldots]{
           \label{wetting-fig}
           Partial wetting of a solid by its own melt is an indirect
           evidence of surface nonmelting.}
\end{figure}

In the language of wetting, SM is nothing else than triple point wetting\cite{dietrich}, 
namely complete wetting of 
the solid-vapor interface by the liquid, in other words wetting of the solid by \emph{its own melt}. 
By contrast, NM corresponds to partial
wetting, schematized in Fig.\,\ref{wetting-fig}.
In this case, there is a clear 
connection between the surface thermodynamic parameters and the angles of partial
wetting \cite{ditolla95} as will be clarified further below.
Thermodynamically, NM will take place whenever the solid surface (solid-vapor
interface) is unable to lower its free energy by turning into a sequence 
of two separate solid-liquid plus liquid-vapor interfaces, namely when
\begin{equation}  \label{eq1}
  \gamma_{\mathrm{SV}} < \gamma_{\mathrm{SL}} + \gamma_{\mathrm{LV}}\,,
\end{equation}
where the $\gamma$'s denote the free energies of the three 
solid-vapor (SV), solid-liquid (SL),
and liquid-vapor (LV) interfaces\,\cite{landau-lifshitz}.
Conversely,
when the inequality (\ref{eq1}) does not hold, SM will ensue instead of NM.
As Fig.\,\ref{pb110-meis-fig} shows, in SM the liquid film, although
technically of divergent
thickness at $T_m$, remains atomically thin until fractions of a degree below
$T_m$.  Interactions propagate readily across such a thin liquid
film. 
The solid-liquid and the liquid-vapor interfaces ``feel''
their mutual presence, so much that they actually merge when $\ell(T)$ is small
enough. For a generic temperature $T<T_m$ the free energy change caused
by melting a solid film into a liquid one of thickness $\ell$ can be written
\begin{equation}  \label{eq2}
 \Delta G/A =
   [(\gamma_{\mathrm{SL}} + \gamma_{\mathrm{LV}}) - \gamma_{\mathrm{SV}}]
   + \ell L \rho_{l}(T_m - T)
   + V(\ell)
\end{equation}
where $A$ is the surface area, $L$ is the latent heat of melting per
unit mass, $\rho_{l}$ is the density of the liquid, and $V(\ell)$ represents a
phenomenological ``interaction'' free energy term between 
the liquid-vapor and the solid-liquid interfaces.
As defined, the interaction $V(\ell)$ vanishes for $\ell \to \infty$, and tends to
$V(0)=\gamma_{\mathrm{SV}} - (\gamma_{\mathrm{SL}} + \gamma_{\mathrm{LV}})$
when $\ell \to 0$, when the dry solid surface free energy is recovered.
Eq.\,(\ref{eq1}) amounts to say that in SM
the interface interaction $V(\ell)$ is globally \emph{repulsive}, whereas in
NM it is \emph{attractive}. A useful mnemonic is that SL-LV repulsion
causes the liquid film to expand (SM), while SL-LV attraction
causes the liquid film to collapse and disappear (NM).
  
The microscopic mechanism behind interface attraction or repulsion varies,
and may be eventually related to such different physical causes as
layering order\,\cite{iarlori96}, van der Waals forces\,\cite{pluis}, or 
molecular order\,\cite{tanya2}. We shall provide specific cases further 
below.

\section{Theory; brief review} \label{sub2}

\subsection{Heuristics: solid surface instability}

Early crude but appealing reasoning leading to SM has been based on heuristics.
Bulk melting usually correlates well with the so called Lindemann melting
criterion: when the r.m.s.\ thermal vibration amplitude of the solid reaches
some 14\% of the nearest-neighbor distance\,\cite{hansen}, 
solids generally melt. Atoms at surfaces
are less coordinated and generally shakier than in bulk, and their vibration 
amplitudes are correspondingly larger. Surfaces will thus reach the 
Lindemann instability vibration amplitude at a lower temperature 
than the bulk. A model embodying the bulk Lindemann criterion is 
the mechanical thermal instability model\,\cite{fukuyama-platzman}. 
An infinite solid will, if prevented from melting, 
become mechanically unstable at a sufficiently high temperature. 
Although the ideal mechanical instability temperature of a solid is
different, and of course somewhat higher than the true melting 
temperature $T_m$ (where the free energy crossing of 
solid and liquid phases takes place), it can nonetheless heuristically 
be taken as a qualitative indicator of the tendency of the solid to melt.
Adopting that line, one can ask whether the stability of the {\em surface} 
will not cease before that of the bulk. In a model semi-infinite solid 
it is found indeed that the mechanical instability of the first surface 
layer occurs at a temperature which is only 73\% that of the bulk\,\cite{pietronero-tosatti}.
This suggests -- correctly -- that surfaces should begin to soften and
melt at about 3/4 the bulk melting temperature. That had been long 
known and was noted by Tammann and Stranski\,\cite{tammann,stranski,nenow}.
Self-consistent surface phonon calculations\,\cite{jayanthi85} and also
experiments\,\cite{frenkenAnharm} later revealed a very pronounced
anharmonic outward expansion of the first surface layer relative to that
of the bulk. Although it is not really possible to generalize, this is
undoubtedly one of the qualitative elements heralding a stronger tendency 
of a surface to become unstable before the bulk.

\subsection{Density functional theory of surface melting}
 
\begin{figure}
\begin{center}
   \includegraphics[width=1.1\textwidth]{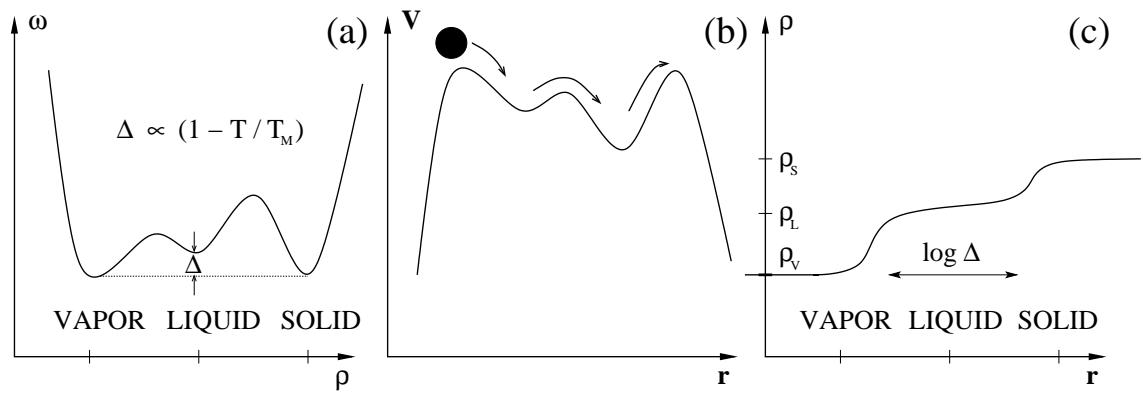}
     \caption[Density functional theory of surface melting\ldots]{
              \label{schemes}
             (a) Minimizing grand potential (Eq.\,\protect\ref{eq4}), 
             (b) Mechanical equivalent (see text), 
	     (c) Density profile, showing that the SV 
	         interface has split into a SL plus LV
                 interfaces.}
\end{center}
\end{figure}

The simplest proper theory of SM is based on optimizing the order parameter
profile, in particular the density profile
$\rho(\mathbf{x})$ of a solid-vapor interface in a system with short range
interactions.
Assuming a bulk grand potential density $\omega = \omega(\rho)$, the solid
vapor coexistence near the triple point as in Fig.\,\ref{pt-plane-fig}
implies two equivalent solid and vapor 
minima of the grand potential per unit volume $\omega(\rho)$
plus one additional secondary liquid minimum with a higher
grand potential density.
The total grand potential of the inhomogeneous system
can be written
\begin{equation}  \label{eq3}
\Omega = \int \!\!\! \int \!\!\! \int d^{3}x \,
    \left[  \omega(\rho(\mathbf{x})) +
            \frac{J}{2} \left| \nabla \rho \right|^{2}
    \right]
\end{equation}
In this Ginzburg-Landau (or Cahn-Hilliard) type phenomenological free energy
form (acceptable for a system with short range forces) the first
term represents the thermodynamic potential density for a uniform system,
while the
gradient
term signifies the extra cost caused by any spatial change in the order
parameter.  For a solid-vapor interface\,\cite{lipowski} we may assume $\rho=\rho(z)$ and
seek to minimize $\Omega$ (per unit area):
\begin{equation}  \label{eq4}
 \Omega/A = \int_{-\infty}^{+\infty} dz \, \left[
            \omega(\rho(z)) + \frac{J}{2} \left(\frac{d\rho}{dz}\right)^2
            \right] = \min
\end{equation}
with the constraints $\rho(-\infty)=\rho_{s}$ and
$\rho(+\infty)=\rho_{v}$, where $\rho_{s}$ and $\rho_{v}$ are solid and vapor 
densities respectively. Amusingly, this minimization problem is formally
identical to the minimum action problem for the one dimensional
lagrangian motion of a classical point object of coordinate $\rho$, mass
$J$, as a function of time $z$, in a potential $-\omega(\rho)$. Starting at
time $z = - \infty$ with zero kinetic 
energy and with a small $\rho =  \rho_{v}$ 
from a first hilltop  of height $-\omega(\rho_{v})$ (the vapor phase), the
point moves
``downhill'' eventually reaching at time $z = + \infty$ the last, and exactly 
equivalent, hilltop $-\omega(\rho_{s})= -\omega(\rho_{v})$ (the solid phase)
(Fig.\,\ref{schemes}).

At temperatures well below the melting point, there is no liquid phase (not even
metastable), meaning that there exists no liquid-like
local minimum in $\omega(\rho)$.  Thus $-\omega(\rho)$ has no
other maxima than the vapor and solid hilltops, and the resulting
solid-vapor interface is unsplit and featureless. Close to $T_m$ however
$-\omega(\rho_{v})$ develops a secondary maximum---a lower hilltop, as it
were---at an intermediate  $\rho=\rho_{l}$. En route
from the vapor to the solid hilltops, the point particle must negotiate this
intermediate ``mountain pass'', where it will not stop, but will still slow 
down considerably. It is easy to check that if $\Delta =
-\omega(\rho_{s})-(-\omega(\rho_{l}))$ 
is the height difference between main and secondary hilltops (the free energy 
difference between solid and liquid, that goes to zero only at the
melting point), 
then the time $\delta z $ the point will spend near the liquid hill, and 
therefore with $\rho \approx \rho_{l}$ is proportional to $\log \Delta$.
Thus the interface density profile is now split
into two interfaces: first a vapor-liquid one, then a liquid-solid one. 
They are separated by a liquid film of density close to $\rho_{l}$, 
whose thickness $\ell = \delta z$ 
is logarithmically increasing as $T \to T_m$. And this is of course just
surface melting.

The above also suggests that one might identify some temperature $T_w < T_m$
where the \emph{bulk} free energy first develops the local minimum
corresponding to the liquid phase, with the temperature $T_w$ where a solid
surface is likely to begin wetting itself with the thinnest liquid film.
It must be underscored that this simple theory only describes SM and
cannot account for NM or for many other complications, upon which we
shall return below. The presence of a ``surface term'' in Eq.(\,\ref{eq4})
will additionally modify this scenario\cite{lipowski}.

\subsection{Lattice mean field theory of surface melting (Trayanov-Tosatti), 
 and beyond}

\begin{figure}
  \begin{center}
  \includegraphics[width=0.8\textwidth]{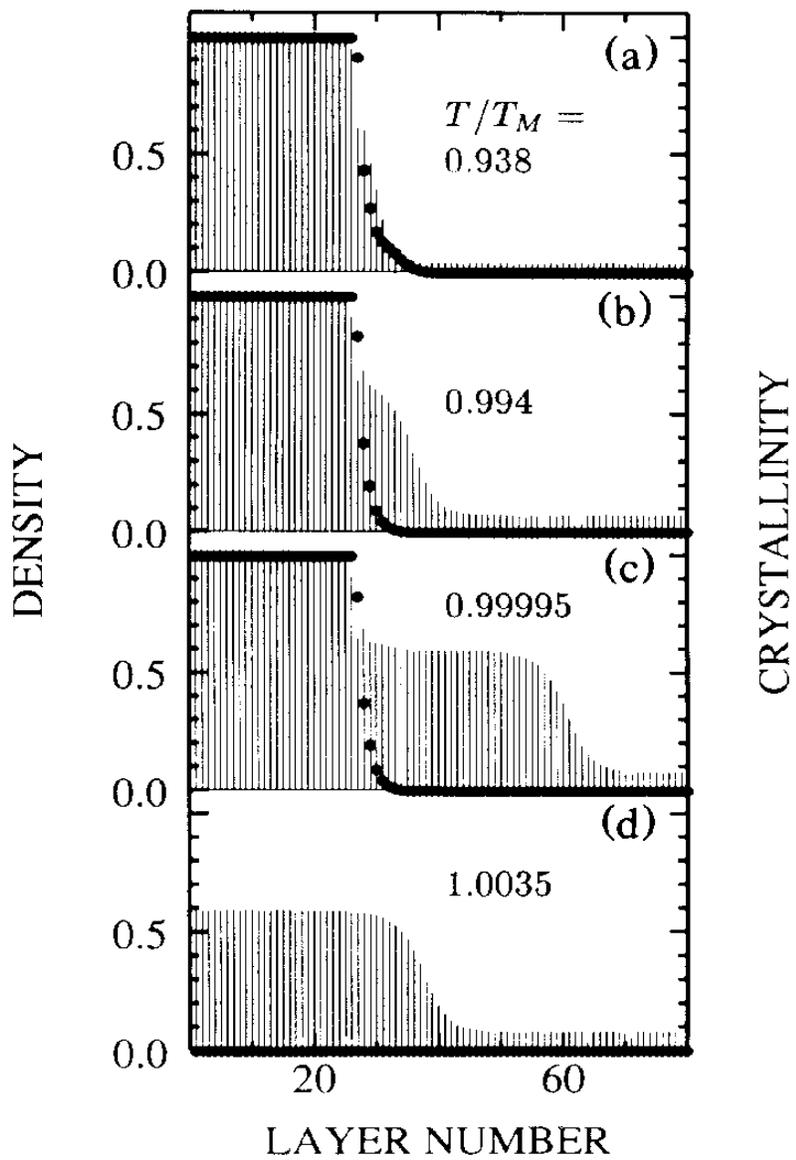}
     \caption[Surface melting of Lennard-Jones(110)\ldots]{
           \label{trayanov-tosatti-fig}
           Surface melting of LJ(110): density profile and cristallinity
           profile. From Ref.\,\protect\cite{trayanov-tosatti}.} 
\end{center}
\end{figure}

The discussion above is centered on density as the sole order parameter.
A solid actually differs from the liquid not just by the average atomic 
density $\rho_0$, but also by the infinite set of crystalline Fourier 
components of the atomic density $\rho_{\bf G}$ ($\bf G$ denoting reciprocal lattice
vectors) that are nonzero only in the solid.
A better order parameter theory of SM than that outlined above should  
be based on a free energy expression that correctly includes these 
crystalline order parameters. Trayanov and Tosatti\,\cite{trayanov-tosatti}
built a simplified but microscopic two-order parameter lattice 
theory based on the average density $\rho_0$  plus a second ``crystallinity'' 
order parameter $c$, the latter effectively replacing the less
manageable infinite set of $\rho_{\bf G}$. Through a further mean field approximation 
where lattice layers are assumed to behave uniformly (thus neglecting roughening
fluctuations), realistic systems such as Lennard-Jones (LJ)
solid surfaces can be described in this manner.
Fig.\,\ref{trayanov-tosatti-fig} illustrates the surface melting of fcc
LJ(110) described by this lattice theory. The approach 
also demonstrated how in the LJ systems the logarithmic liquid film thickness 
divergence close to $T_m$ is asymptotically replaced by a more realistic
$(T_m - T)^{-1/3}$ divergence, due to the (omnipresent) van der Waals long range
interatomic potential tails. Let us expand briefly on this point.
Given three media S, L and V, with a finite thickness $\ell$ of L sandwiched
between semi-infinite S and V, there will generally arise a long range
tail to the interaction free energy between the two interfaces 
\begin{equation} \label{eqhamak}
  V(l) \sim  H/\ell^2 
\end{equation}
due to long range
dispersion forces. The so-called Hamaker constant $H$ is generally positive
when the intermediate L phase is dielectrically less dense that S, and more
dense than V\,\cite{israelachvili}. Positive $H$  means long range
repulsion between the interfaces, which favors SM. That is precisely the case
for the LJ-like systems, where the liquid is less dense than the
solid. The same will generally also apply to many metals and other materials. 
In different cases, such as valence semiconductors and semimetals, melting leads 
to a denser and generally metallic liquid state, and there $H<0$. Negative expansion
at melting and a negative Hamaker constant occur also in other notable 
cases, water among them\,\cite{dash,dash1}. Here $H<0$ means a long range attraction between 
interfaces, and that implies NM. As we shall see below,
the surfaces of a semiconductor actually appear to wet close to $T_m$. 
However the wetting is incomplete, and the melted film thickness $\ell$ remains microscopically
thin, in agreement with their negative Hamaker constant.
 
Of course mean field theories, such as Trayanov-Tosatti above, are only
an approximation to a real critical phenomenon. Mean field is 
usually a bad approximation sufficiently close to a critical point 
(in SM the critical point is $T=T_m$), where the correlation length 
(here signified by the liquid thickness $\ell$)
diverges.  In systems with short range forces critical fluctuations may
lead to nonclassical exponents. For SM, this aspect was discussed
theoretically by Lipowsky and collaborators\,\cite{lipowski}, and later 
by Chernov and Mikheev\,\cite{chernov-mikheev}. It was also addressed 
through atomistic simulations by Chen \emph{et al}\,\cite{chen}. The 
conclusion is that although non-mean field fluctuations (mostly roughening fluctuations)
actually have a large quantitative effect\,\cite{chen}, they will not 
change the mean field exponent $-1/3$ which originates from long range forces\,\cite{pluis}.

\subsection{Simulations: Molecular Dynamics}

\begin{figure}
\begin{center}
   \includegraphics[width=\textwidth]{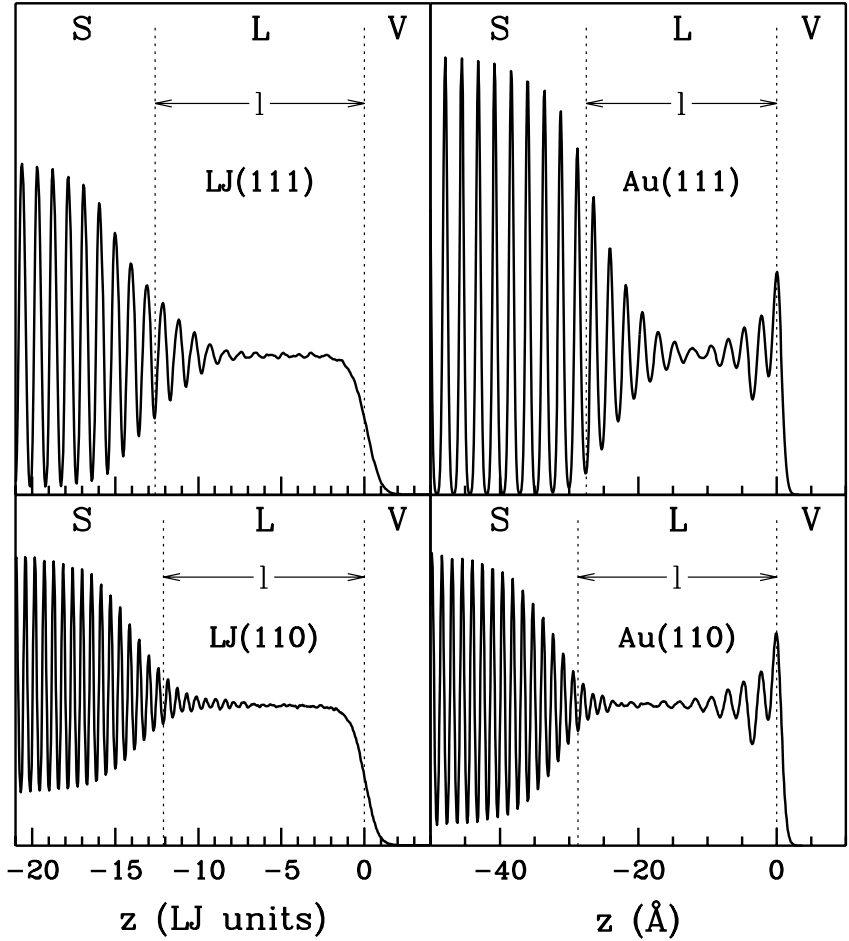}
     \caption[Typical solid-liquid-vapor density profiles\ldots]{
            \label{ljau}
            Typical solid-liquid-vapor density profiles
            obtained by simulations
            of Lennard-Jones(110), (111) and Au(110)  
            (three surfaces that undergo SM) just below $T_m$. For 
            Au(111) (which is instead a NM surface) a nonequilibrium
            configuration is shown with two facing equal-period density 
	    oscillations which lead to 
            attraction and eventually to the collapse
            of the two interfaces (see text).
            From Ref.\,\protect\cite{ditollabook}.}  
 \end{center}
 \end{figure}

Atomistic simulations constitute a very important tool in the field 
of melting and surface melting. They can be either of Monte 
Carlo (MC) type, or of
Molecular Dynamics (MD) type. We shall concentrate 
on the latter, using work done in our group for specific examples.
SM simulations are generally conducted on crystal slabs,
made up of a sufficiently large number of atomic solid layers, with two free
surfaces, or with one free and one ``frozen'' (fixed atoms) surface, and periodic boundary conditions along the two directions
parallel to the surfaces. The in-plane 
simulation cell size must be allowed or otherwise made to expand gradually 
as temperature increases, to guarantee that the planar stress in the solid 
on account of thermal
expansion remains as close as possible to zero. 
The simulation consists of solving Newton's equations of motion
\begin{equation}
 M_i \ddot{\bf R}_i  = - {\bf \nabla}_{i} E_{\mathrm{pot}}(\{{\bf R}\})
\end{equation}
for all particles of mass $M_i$, and coordinates ${\bf R}_i$. The basic input
needed is the potential energy $E_{\mathrm{pot}}(\{{\bf R}\})$ as a function
of all coordinates. 
Depending on systems, and on the accuracy needed, the choice is between
empirical interatomic potentials on one side, and first principles total
energy calculations $E_{\mathrm{pot}}(\{{\bf R}\})$ on the other side.
Empirical two-body potentials, such as Lennard-Jones (LJ), are quite
reasonable for rare gas solids or van der Waals molecular crystals,
but not for most other solids. For metals, empirical many body
potentials such as the Embedded Atom Model\,\cite{dawbaskes}, the 
similar glue model\,\cite{ercolessi-parrinello-tosatti88}, the
Finnis-Sinclair potential\,\cite{finnis-sinclair} etc., are much more
suitable than two-body forces. Many-body empirical potentials
were developed also for such systems as valence semiconductors 
\cite{brenner,tersoff} but here the first principles simulation 
approach is generally much more appropriate\,\cite{takeuchi}.
Typical 
SM simulation density profile outputs are shown in Fig.\,\ref{ljau} 
for a modified LJ system and for a Au metal surface respectively.
Technical details and some further results for simulations of SM can be
found in the review article by Di Tolla {\em et al}\,\ \cite{ditollabook}. 
Simulations may in many ways replace experiment, and must be similarly
regarded and understood by means of theory---not discounted as
self-explanatory as it is sometimes tempting to do. Similar to experiment, they do
uncover (occasionally by serendipity), novel or unexpected behavior. 
Such has been the case for surface nonmelting, first discovered in
simulation\,\cite{carnevali} and next found experimentally\,\cite{frenken87},
and several other phenomena, to be discussed further below.

\section{Surface melting and related phenomena in real solids}

\subsection{van der Waals solids: interplay of roughening, preroughening,
and melting}

Rare gas solids are reasonably described through a Lennard-Jones
potential.  As first shown by extensive MD studies by Broughton and
collaborators\,\cite{broughton,broughton1} they are expected to display surface melting on
all crystalline faces. Here truly microscopic experiments are not
abundant, because of technical difficulties.  There is however a large body of
work by the surface adsorption community, interested in understanding
adsorption isotherms as a function of temperature\,\cite{taub92}. 

Adsorbtion of atoms on a substrate generally takes place in a
layer-by-layer mode, up to the so-called surface roughening
temperature $T_R$, where the interface width diverges and layers cease
to exist. Surface roughening--a Kosterlitz-Thouless phase transition
taking place (sometimes but not always) at a temperature $T_R < T_m$---is 
a totally distinguished phenomenon from surface melting\,\cite{bernasconi}, 
but it does have some relations with it. It can be argued for example
that surfaces that undergo SM at $T=T_m$ will necessarily undergo a
roughening transition at $T_R < T_m$\,\cite{levi-tosatti86}. That is 
because a liquid surface is technically rough, so that capillary 
fluctuations cause the liquid-vapor interface width to diverge. 
Conversely, although there is apparently no theorem that says so, a NM
surface will usually not undergo roughening below $T_m$. Disregarding
connections with surface melting---even though they are sometimes 
important, as discussed below---roughening is generally discussed
in purely lattice models.

\begin{figure}
 \newpage
 \begin{center}
   \includegraphics[width=0.8\textwidth]{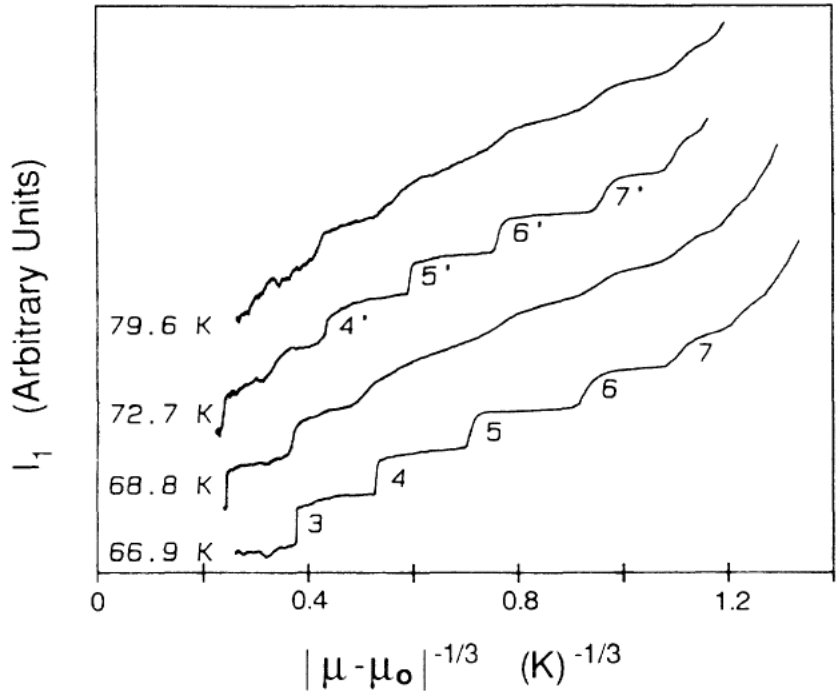}
   \caption[Reentrant layering during the deposition\ldots]{
           \label{hess-fig}
           Reentrant layering during the deposition of Argon on a graphite
           substrate. On the increasing $T$, the ellipsometric coverage isoterms switch from a
           staircase profile (growth layer-by-layer) to a continous one at 68.8K,
           and then again to a staircase profile as temperature is
           further increased. The new structure consists of half layers, instead of full layers.
	   Image and data from \protect\cite{hess}.}
  \end{center}
 \end{figure}

\begin{figure}
\newpage
 \begin{center}
 \includegraphics[width=0.6\textwidth]{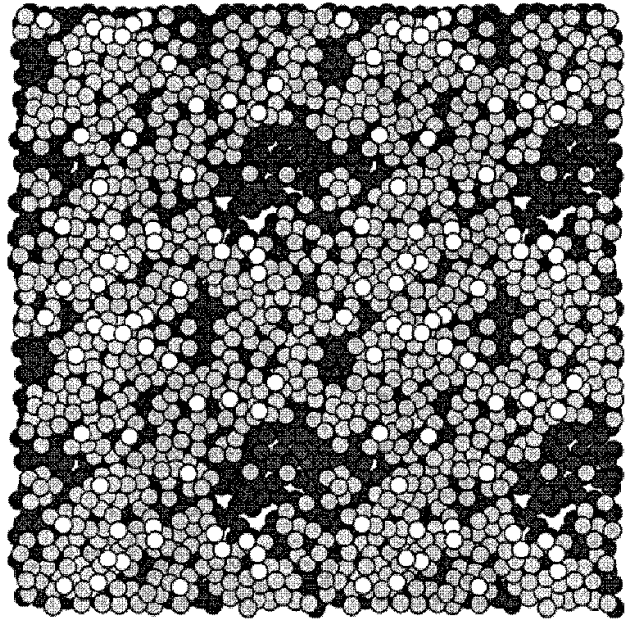}
 \caption[Top view of the Ar(111)\ldots]{
          \label{celestini-fig}
             Top view (snapshot) of the three outer layers of simulated grand canonical
             Ar(111) at $0.8\,T_m$. This surface is nearly ``disordered flat''. Atoms
             are represented as
             black, grey or white if their $z$ coordinate indicates that they belong respectively 
             to the subsurface layer, the surface layer or the adatom layer.
             For convenience, four adjacent 
             simulation cells are shown.  The nearly half occupancy of the surface layer (layer 1) 
             is realized through large islands and large craters. Image from Ref.\,\protect\cite{celestini0}.}
\end{center}
\end{figure}

There is yet another surface phase transition that may occur below
melting, in fact below roughening, therefore called
preroughening (PR)\,\cite{dennijs,dennijs1,santoro,santoro1}. PR is associated with the bulk
lattice planes having some stacking such as $abab\ldots$ or $abcabc\ldots$ where
planes $a$ and $b$ can only be made to coincide through a fractional
translation. At low temperatures a perfect surface, say a (100) face, will 
with identical probability be either $a$ or $b$ terminated. As in an Ising
model, either choice will ``spontaneously break'' the symmetry between 
$a$ and $b$ . At the PR temperature $T_{PR}$, a surface restores 
precisely the $a$-$b$ symmetry in an Ising-like fashion. 
To achieve that, the topmost atomic layer undergoes a most surprising 
spontaneous rearrangement, from full coverage below $T_{PR}$,
 (full $a$ or full $b$) to 
$\frac{1}{2}$ coverage above $T_{PR}$. The excess atoms simply migrate
away, turning the top half layer into a sort of soft a-b ``checkerboard''. 
Strictly at $T=T_{PR}$ the surface becomes technically rough, only to 
become flat again between $T_{PR}$ and $T_R$. 
Preroughening was first discovered
theoretically in lattice models\,\cite{dennijs}, and received little 
attention by experimentalists until the equally surprising phenomenon of 
re-entrant layering was discovered\,\cite{hess,hess1}.
Adsorbtion of Ar, Xe, Kr on a graphite substrate
(Fig.\,\ref{hess-fig}) proceeded by full layers below $0.82\,T_{m}$. This
appeared to be roughening: but layering surprisingly re-entered 
at higher temperature, this time by \emph{half}\ layers.
While that clearly called for an explanation in terms of
preroughening \,\cite{dennijs91} a considerable controversy arose, as
independent evidence suggested that the first surface layer was in reality no
longer solid, with considerable liquid-like disorder and first
layer mobility in that temperature regime\,\cite{larese}. 
Moreover, reentrant layering signaled a clear first order transition, 
whereas PR should be second order. The resolution of this puzzle, reached through
theory\,\cite{jagla98}, grand canonical MC\,\cite{celestini0,celestini00} and MD\,\cite{jayanthi00} simulations, 
proved to be that
PR is linked to first (half-) layer surface melting: in essence
both phenomena take place simultaneously. PR in a rigid lattice 
surface model would indeed only take place at higher temperatures, and
the half monolayer would in that case be stabilized through a second order transition.
In real off-lattice Argon however the half monolayer soft checkerboard
immediately melts, with a first order transition and a large positional
entropy gain over the corresponding hypothetical solid half monolayer. 
As a result, preroughening and first layer melting actually take place 
simultaneously, at a much lower temperature than either of them separately, 
and of course through a first order transition. Fig.\,\ref{celestini-fig}
shows a 
realistic snapshot picture, obtained by MC, of the Ar(111) surface 
simulated in the reentrant layering temperature regime, showing the spontaneous
formation of a roughly half-coverage, liquid like top monolayer.

\subsection{Ionic insulators}

Surfaces of ionic crystals such as NaCl, other alkali halides, MgO, etc, 
are often used as substrates for growing other materials, and their properties
are therefore better known far below the melting point. Yet, some data 
are also available about their behavior at $T_m$. 

\begin{figure}
 \begin{center}
  \includegraphics[width=0.6\textwidth]{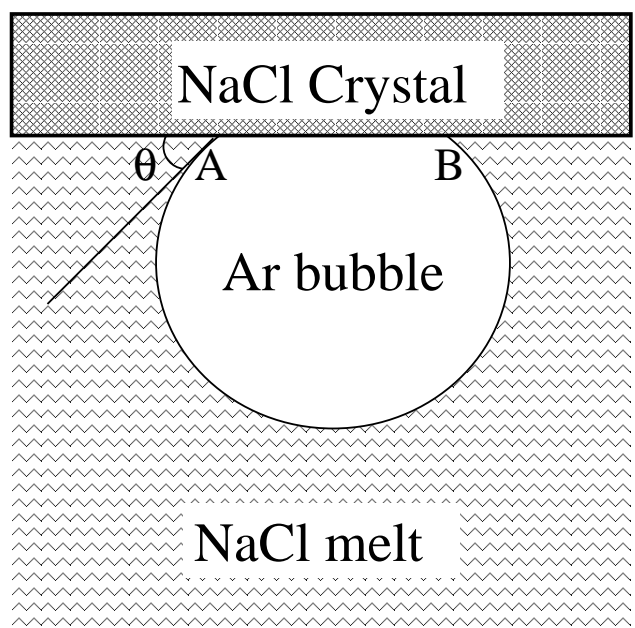}
  \caption[Argon bubble in contact with the solid\ldots]{
           \label{NaClexp}
           Argon bubble studies of liquid NaCl in contact with the solid. 
	   The solid-liquid-vapor junctions A and B reveal 
           a surprising lack of complete wetting, with a large partial wetting
           angle of about 48 degrees.} 
 \end{center}
\end{figure}

\begin{figure}
  \includegraphics[width=1.1\textwidth]{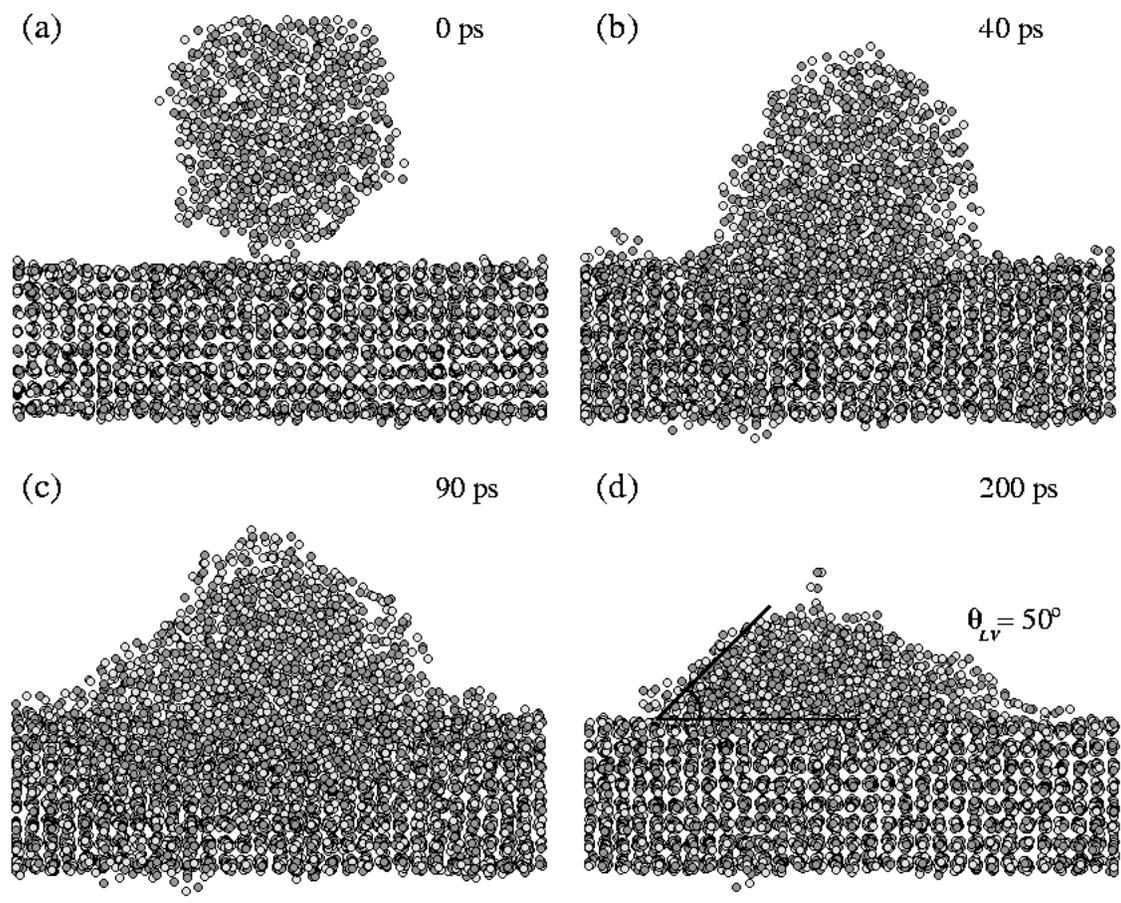}
  \caption[Time evolution of NaCl liquid nanodroplet\ldots]{
            \label{NaClsim}
            The simulated time evolution of NaCl liquid nanodroplet brought into the contact with NaCl(100)
            at the melting point. After 100 ps the drop stabilizes in the metastable 
	    state forming a partial wetting contact angle $\theta = ( 50 \pm 5)^{\circ}$
	    From Ref.\,\protect\cite{tanya1}.} 
\end{figure}

In this respect, NaCl appears to be one of the best studied, and can therefore
be chosen as a case study. Argon bubble studies of liquid NaCl in
contact with the solid revealed a surprising lack of complete wetting,
with a large partial wetting angle of about 48
degrees (Fig.\,\ref{NaClexp})\cite{mutaftschiev} .
Very recent MD simulations of solid
NaCl(100) clearly demonstrates NM of this solid surface, also predicted to
survive in a metastable state well above $T_m$\,\cite{tanya2}.
Simulations of a droplet of melted NaCl brought into contact with solid
NaCl(100) at $T=T_m$ (Fig.\,\ref{NaClsim}) clearly shows incomplete wetting, 
with
an external partial wetting angle of $(50 \pm 5)^{\circ}$, in good
agreement with the bubble experiments\,\cite{tanya1}.

The microscopic
reasons for the exceptional stability of the solid NaCl(100) surface and its
unreadiness to wet itself are presently under scrutiny. They are intriguing 
in view of a large positive Hamaker constant -- NaCl undergoing a 26\% volume 
expansion at melting -- which would suggest the opposite behavior, namely 
SM. A second element wrongly suggesting SM against NM is the structure 
of the liquid surface, very strongly fluctuating and totally devoid of 
layering (Fig.\,\ref{NaClliq}) unlike e.g., nonmelting metal surfaces 
(see below). So why does strong NM arise in alkali halides?

Preliminary useful clues come from surface thermodynamics. MD simulations 
permit the explicit calculation of $\gamma_{LV}$,$\gamma_{SL}$ and $\gamma_{SV}$ as a
function of $T$\,\cite{tanya2}. It appears that while $\gamma_{SV}$ drops 
very strongly in virtue of large surface relaxations and the associated 
growing vibrational entropy at $T \sim T_m$, the liquid surface is 
incapable of developing a competing amount of entropy, presumably because 
of the all-important local charge neutrality constraint, and of
the related appearance of molecular order in the fluctuating liquid
NaCl surface. This physical result is the subject of current work\,\cite{tanya3}.

\subsection{Valence semiconductors and semimetals}

Valence semiconductors such as Si, Ge, GaAs and semimetals like Ga are
known to turn fully metallic and denser either at high pressure or when
they melt at high temperature.  This is illustrated by the schematic Ga
phase diagram of Fig.\,\ref{ga-fig}. The higher liquid density here causes the Hamaker
constant to be negative\,\cite{chen}, and this 
necessarily hinders complete surface melting. 
Reports of regular SM with
unlimited growth of the liquid film at $T_m$ that have appeared for 
Ga\,\cite{gruetter-bilgram} are incompatible with the negative Hamaker
constant, and might in our view be an artifact, possibly 
due to strains (see below). Due to the long range attraction, regular 
and complete SM is excluded for all faces of these $H<0$ materials.

The clean surfaces of semiconductors,  Ge(111) and Si(111), were
thoroughly investigated for surface melting. While they were found to
disorder\,\cite{vanderveen} and simultaneously to become
metallic\,\cite{modesti-sancrotti} on the topmost layer at some $T<T_m$,
there is clear evidence that the disordered film does not grow as
$T_m$ is approached.  In order to understand theoretically the top layer
disordering and metallization, MD simulations are perhaps the best tool.
Since electronic effects and the semiconductor-metal transition play 
a very important role in the melting of these systems, it seems mandatory 
to take recourse in this case to first-principles simulations, where forces 
acting on the ions are obtained by solving the full electronic 
structure problem. Simulations of this type, pioneered by Car and
Parrinello in the mid 80s\,\cite{car-parrinello}, are by their nature 
much more expensive and demanding than the empirical ones. Results for 
Ge(111) by Takeuchi {\em et al}\,\,\cite{takeuchi} are shown in Fig.\,\ref{takeuchi-fig} at $T \sim T_m$.
As signaled by the broader density profile, the topmost bilayer is indeed 
strongly disordered, but disorder does not propagate to the second bilayer, 
indicating incomplete melting. Electronically, the surface disordered film 
is metallic\,\cite{modesti-sancrotti}, as opposed to the underlying bulk, which remains semiconducting.

\subsection{Metals}

The surface melting behavior of metallic surfaces has received by
comparison much more experimental attention.  Metal surfaces usually 
possess generally positive Hamaker surfaces, and generally melt. In detail,
they may exhibit either melting and nonmelting, depending on the metal
and on the crystallographic orientation. A minority of close packed
surfaces such as Pb(111)\,\cite{frenken87}, Al(111) and
Al(100)\,\cite{molenbroek}, etc. display NM, and remain in fact smooth and
dry all the way to $T_m$. The vast majority of metal faces consisting of
all other orientations, where packing is poorer, undergo SM. In
intermediate packing cases like for instance Pb(100)\,\cite{frenken87},
the first few layers melt, but the wetting does not proceed and the
liquid film growth is blocked to a finite thickness until $T_m$
(incomplete SM). Although this incomplete melting will macroscopically 
appear just the same as NM---namely partial wetting with a finite wetting
angle---there is a clear microscopic difference because the surface is no
longer dry below $T_m$.

An early review of SM on metals can be found elsewhere\,\cite{vanderveen}. We will limit
ourselves to mention here a few high-temperature metal surface phenomena,
particularly those connected with NM, which have been especially
highlighted in our group.

The first is the existence in surface NM of a critical liquid nucleation
thickness $\ell_{crit}$ which is finite above $T_m$ and only vanishes at some
``surface spinodal temperature'' $T_s > T_m$. This fact, first
discovered in simulation\,\cite{carnevali}, indicates that in NM the solid
surface is in fact metastable, and protected by a nucleation barrier,
between $T_m$ and $T_s$. An immediate consequence is that NM solid
surfaces can be overheated, of course only for a short time, up to ati
most $T_s > T_m$. For the (111) surfaces of Pb, Al and Au the calculated
amount of maximum theoretical surface overheating $T_s - T_m$ is
120\,K, 150\,K, and $\sim$ 150\,K respectively. This possibility was
demonstrated experimentally by laser heating techniques\,\cite{ali}, and
also in small Pb clusters\,\cite{metois}.
  
Another phenomenon discovered theoretically is non-melting induced
surface faceting. Consider a general crystal surface, whose orientation
is close to, but not exactly coincident with, a flat NM face. Such a
``vicinal surface'' will consist of a sequence of flat terraces separated by
surface steps.  At low temperature the steps will usually repel
mutually and form an ordered array, so that the solid vicinal surface is
stable. As $T_m$ is approached, the steps suddenly coalesce in bunches
giving way to much larger flat terraces, separated by very inclined
facets where the steps have bunched up. The inclined facets with high step 
density actually melt, while the step-free flat facets remain 
dry. This non-melting induced surface faceting was first predicted 
thermodynamically\,\cite{nozieres}, as seen on the free energy sketch (Fig.\,\ref{nozieres-fig}), 
demonstrated by MD simulation (Fig.\,\ref{bilal-fig})\,\cite{bilalbegovic} ,
and eventually observed experimentally (Fig.\,\ref{frenken-fig})\,\cite{frenkenfacet} .
It is most likely the reason why e.g., beautifully perfect flat faces 
can be generated in gold, by simple heating close to the melting point. 
Au(111) being a nonmelting face, it will at high temperature cause all 
steps and imperfections to bunch up in some place, sweeping itself clean and flat.

A third interesting point is the microscopic investigation and
clarification of the relationship between solid surface NM and its
partial wetting by a drop of melt. While macroscopic partial wetting with
a nonzero contact angle of the liquid with its own solid is not commonly 
reported for metals, it is very clearly found by
simulation. Fig.\,\ref{ditolladrops-fig} shows the contrasting fate of a
simulated Al liquid nanodroplet at $T=T_m$ when approaching alternatively a SM
Al(110) surface, or a NM Al(111) surface. In the first case, the droplet
spreads rapidly and disappears. In the second it settles down to a
metastable state, forming the nanoscopic equivalent of a partial wetting
angle $\theta_{LV} = (22 \pm 3)^{\circ}$.

There is in fact, in a simple model, a direct relationship
between $\theta_{LV}$ and the overheating $O = T_s - T_m$ of a NM surface,
namely
\begin{equation} \label{eq5}
 \frac{T_s}{T_m} = 1 + \frac{2 \gamma_{LV}}{L \rho_{l} \xi} \sin^{2}{\frac{\theta_{LV}}{2}}   
\end{equation}
where $L$ is the latent heat of melting per unit mass, $\rho_{l}$ is the density 
of the liquid, $\xi$ is the correlation length in the liquid. The background of
this formula is Eq.~(\ref{eq2}) with a short range interface interaction
\begin{equation} \label{eq6}
V(\ell)= V(0) \exp -(\ell/\xi)
\end{equation}
plus the appropriate Young equation\,\cite{ditolla95,nozieres}
\begin{equation} \label{eq7}
\gamma_{SV} = \gamma_{LV} \cos \theta_{LV} + \gamma_{SL} \cos \theta_{SL}
\end{equation}
where moreover $\theta_{SL}\sim 0$ at $T=T_m$.
Based on Eq.~(\ref{eq5}) the calculated ideal overheating temperatures of 
Pb(111), Al(111) and Au(111) given earlier above predict partial self-wetting 
angles $\theta_{LV}$ of
$(16\pm 1)^{\circ}$, $(18 \pm 2)^{\circ}$, $(33 \pm 2)^{\circ}$ respectively.
By comparison $\theta_{LV}$ is found to be $(14\pm 1)^{\circ}$\,\cite{frenkenfacet}
in the experiment for Pb(111), and $(22 \pm 2)^{\circ}$ in the simulation for Al(111)
(see above).

In addition to the above thermodynamics, there is 
a simple microscopic understanding of the physics that leads to
NM of these close packed metal surfaces\,\cite{ditolla95,ditollabook}. It has to do with the intrinsic atomic 
structure of the two SL and LV interfaces involved. A series of $z$-resolved 
in-plane averaged density profiles of the SL and LV interfaces as displayed 
by MD simulations of suitably prepared liquid films on Au(111), Au(110), 
and LJ(111) near their respective $T=T_m$ is shown in Fig.\,\ref{ljau}. 
A well defined damped density oscillation is seen to propagate 
from the solid into the liquid, carrying approximately the solid interplanar 
distance as a wavelength. This wavelength of course depends on the
crystallographic direction; in the present example it is large for (111)
but small for (110). 
A second damped density oscillation starts at the liquid surface,
carrying {\em inward} a generally different wavelength, determined this
time by the main peak in the liquid structure factor $S(k)$. This second
oscillation, essentially non-existent in the LJ liquid, is generally
strong in a metal (see discussion further below). Being a property of the liquid,
the surface layering oscillation of the surface melted film is obviously
face independent. 

When, as in Au(110), 
the solid and liquid oscillations facing one another 
possess wavelengths that are out of tune, then their superposition is unfavorable 
and causes interface repulsion, eventually leading to SM. This is what happens 
for Au(110). When instead, as in Au(111), the two oscillations are close to 
being perfectly tuned, their superposition is favorable and causes interface 
attraction, leading to NM. Finally, the practical absence of a layering oscillation 
in the LJ liquid surface indicates indifference of the two interfaces. They nonetheless 
eventually interact via the positive Hamaker constant [Eq.~(\ref{eqhamak})], and 
therefore SM ensues in that case too.

\section{Special cases}

\subsection{Strained solids}
In an infinite bulk solid, a shift $\Delta P$ of external pressure 
provokes a shift of the bulk melting temperature by an amount 
$\Delta P (v_{l}-v_{s})/(s_{l}-s_{s})$, where $v_{l}$
and $v_{s}$ are the molar volumes of the liquid and the solid phase
respectively, and $s_{l}$ and $s_{s}$ are the corresponding molar
entropies. Thus an increase of the external pressure generally
increases the melting temperature. Contrary to this, a uniaxial
strain always works the other way around and favors melting, irrespective
of the sign of strain. Uniaxial compression or stretching
increases the elastic energy of the solid (so long as the solid
can sustain the corresponding stress without relevant plastic
deformation). On the other hand the liquid does not support
shear, and all elastic energy is released by flow. The bulk melting 
temperature of a strained solid is thus lowered by an amount 
which is quadratic in the anisotropic strain. This effect has been 
experimentally observed in He crystals\,\cite{balibar}. In metals it
has been studied theoretically through simple thermodynamics, and 
accessed directly by molecular dynamics simulations for aluminum\,\cite{ugo,ugo1}.
This is therefore really a bulk effect though it will readily show 
up at surfaces, as shown by the simulation
results of Fig.\,\ref{meltingstrain-fig}. Due to the strain-induced lowering of the
bulk $T_m$
the liquid film thickness at fixed temperature on Al(110) increases 
strongly and quadratically with strain, due simply to strain induced 
lowering of $T_m$. 
A corollary of some practical importance is that in any study of surface
melting the strain conditions must be severely controlled. The Al test 
case indicates that a relatively small strain of $10^{-3}$ implies
a shift of $T_m$ of 0.03 degrees, or 0.003\%.
As such a degree of precision is sometimes approached in surface melting
studies, the error introduced by strains can be significant.
Since strain 
conditions are seldom specified explicitly, it seems possible and in some cases likely 
that some of the asymptotic SM data in the literature might need re-checking 
against that source of error.
  
Besides this bulk effect, there are more interesting strain-related
surface effects. The first is the so-called Asaro-Tiller-Grinfeld
instability, an experimentally well established effect; the second,
is a strain-induced prewetting and eventual NM of an initially SM 
surface, an effect so far only hypothetical, but observed in MD simulation. 

The Asaro-Tiller-Grinfeld instability\,\cite{asaro-tiller,grinfeld} is 
a periodic wavy modulation of the interface between the liquid and the 
strained solid, once the in-plane strain exceeds a critical threshold. It is connected 
with the possibility, permitted by a wavy interface only, of the solid 
``promontories'' sticking out into the liquid to relax, reducing their strain 
and the associated elastic energy. When the bulk strain magnitude is large,
the local surface strain release becomes large enough, and pays for 
the free energy cost associated with the extra interface area due 
to the waviness. This effect has been observed at the He solid-liquid 
interface by the Ecole Normale group\,\cite{balibar}.

Strain induced prewetting can instead occur because a SV interface will
generally possess, in full equilibrium, a nonvanishing surface stress. 
For instance most metal surfaces are known to possess a large tensile
surface stress, signifying that the topmost layers would really like 
to contract relative the bulk, whose lattice spacing is of course
fixed\,\cite{ibach}.
When an in-plane bulk strain is applied, surface stress will work 
against the strain, or along with the strain, and the work reversibly 
done or released will increase or decrease respectively the overall SV surface 
free energy $\gamma_{SV}$. As seen in Eq(\ref{eq1}) and (\ref{eq2}) a
strain-induced increase 
of $\gamma_{SV}$ will further encourage SM, but a {\em decrease} will 
oppose it. In the latter case it can be shown (Fig.\,\ref{diagram}) in the very 
simple model
of ref.~\cite{ugo} that strain will cause ordinary, continuous SM to be replaced 
by a {\em prewetting transition}: as $T$ grows
surface melting ceases to be continuous, and develops a sudden jump in the
liquid film thickness $\ell$ from zero to a finite value, from where it then grows 
continuously and diverges at $T \to T_m$ as in regular SM. This 
strain-induced prewetting transition has recently been confirmed by 
simulations\,\cite{ugo1}, but has yet to be experimentally pursued. 

\subsection{Defreezing of glass surfaces}

We dealt thus far with surfaces that are not just solid, but also crystalline.
Many solids are however not crystalline, but amorphous or glassy.  
A glass does not break translational invariance the way a crystal does.
Still, it breaks ergodicity in exactly the same manner as the crystal. Due
to freezing and to the ensuing dynamical arrest, at the bulk glass 
temperature $T_g$ a glass, similar to a crystal, ceases to explore the 
full configurational phase space in the extensive way a liquid does, 
and acquires {\em rigidity}\ \,\cite{anderson}, the hallmark of broken positional 
ergodicity.

Let us consider a semi-infinite glass with a free surface, and let $T$ increase:
will the glass surface defreeze at the same temperature $T_g$ as the bulk glass? In
principle there is no simple answer to that, because the transition to the
glassy state is a matter of dynamics, and not of thermodynamics. One can
nonetheless build a simple ``enthalpy functional approach'', different in
spirit but relatively parallel in reasoning, to the density functional
theory of SM of Sec.\,\ref{sub2}. This approach is in turn based on a recent
thermodynamic formulation of the glass transition, where this transition
is in fact second order\,\cite{speedy,speedy1}. 

The enthalpy functional analog of Eq.\,\ref{eq4}
\,\cite{jagla-tosatti1} which represents the surface version of the 
thermodynamical glass theory indeed predicts  a {\em surface defreezing} 
beginning at a lower temperature than the bulk $T_g$, with a penetration 
depth that also grows logarithmically as $T \to T_g$, superficially similar 
to SM in crystal surfaces. At a closer look, surface glass defreezing is in reality a much 
weaker affair than crystal SM is. To illustrate this, one can note a 
very substantial difference in the amount of total surface diffusivity, 
given by the integral of surface diffusivity over depth. This integral, 
measuring the amount of real flowing liquid at the surface, is divergent 
at $T_m$ for a SM crystal surface, but remains finite up to $T_g$ at a 
glass surface.

Glasses may be expected to flow very, very slowly.
If the flow takes place in the bulk, a piece of glass with a given initial
shape will generally change its shape in a well defined manner\,\cite{stokes}.
The expected shape change will be totally different if the flow takes place
{\em at the surface} instead of the bulk\,\cite{jagla-tosatti2}. Unlike bulk flow,
surface flow should wipe out sharp corners and edges, rounding off a finite
but increasing length away from the corner as a function of time. It would be
very interesting if these predictions, so far purely theoretical, could
be explored experimentally, e.g., by studying shape changes of glasses by 
centrifuge experiments.

\subsection{Rotational surface melting in molecular solids}

We have been concerned with surfaces consisting implicitly of either atomic 
elements, or of small molecules such as NaCl merging together to form the solid.
Let us consider by contrast molecular crystals made up of molecules 
that do not merge when crystallizing, and behave so to speak as nanosized 
rigid bodies, both in the crystal and in the liquid phase. In that case thermal 
disordering will concern, besides the positional degrees of freedom of the 
molecular centers of mass, also the {\em rotational} degrees of freedom. 
For very asymmetrical molecules,
it can be expected that the two sets of degrees of freedom will be closely
entangled. For nearly spherical molecules on the other hand the rotational 
and positional phenomena will involve very different energies
and therefore different temperatures, the rotational well below
the positional. In that situation one can have rotational
melting, a first order phase transition transforming a proper crystal into 
a ``plastic crystal'' (somewhat of a misnomer) where the molecular centers
of mass are still positionally crystalline, but where the rotational
coordinates are thermally disordered, or melted.
A particularly popular example of this behavior is provided by $C_{60}$ fullerite,
where rotational melting is known to take place at $T_{rot} \sim 250-260$\,K\,\cite{heiney}, 
whereas there is in $C_{60}$
no positional melting, and essentially no liquid phase. 

Consider the surface of such a molecular crystal and let $T$ increase towards $T_{rot}$.
Will there be a rotational equivalent to surface melting, the first few
layers thermally disordered and the bulk still rotationally ordered?
The answer is in principle yes; and again, simple free energy functionals
can be resorted to in order to explore that scenario\,\cite{passerone-tosatti}.
Here, perhaps the most interesting aspect here, (and certainly the most
relevant to $C_{60}$, where an abrupt surface phase transition was detected some 30 degrees
below  $T_{rot}$ \,\cite{modesti}) appears to be the earliest stages of surface
rotational disordering. 

There is a crucial difference between rotational and positional melting. 
Even in a rotationally melted plastic crystal,
a molecule is still surrounded by a lattice, its cage exerting onto the molecule 
a sort of crystal field potential. In the ordinary liquid instead there is no 
remaining lattice at all. In the molecular solid, the crystal field potential felt 
by a first layer 
surface molecule is very different from the bulk one. The early stages of 
surface disordering and their precise nature will be vastly determined 
by that difference. Calculations
show that at the surface of $C_{60}$ the crystal field actually {\em frustrates} bulk
order; moreover it does so differently for different surface molecules. This gives
in fact rise to a first layer thermal disordering process that proceeds in two stages,
the more frustrated molecules disordering at a lower temperature, the less frustrated
at higher temperature (but still below the bulk $T_{rot}$)\,\cite{laforge}.

Similar scenarios could be expected to take place for other molecular crystals, 
where surface studies are however not yet abundant so far.

\section{Liquid metal surfaces}

Our focus throughout has been on solid surfaces, touching only incidentally upon
the physics of liquid surfaces, whenever that was important for surface
melting. Yet, liquid surfaces have so to speak a life of their own. Liquid metal 
surfaces were long known to be interesting and structured. As revealed by e.g., X-ray
reflectivity experiments\,\cite{pershan,pershan1}, and as discussed long ago by chemical physicists
such as Rice\,\cite{rice}, the density profile of liquid metal surfaces 
shows a tendency to layering. Figs.\,\ref{pb-fig} and \ref{lj-fig} show as an example the density profiles
obtained by simulation for the Pb, Al, LJ and NaCl liquid surfaces at their respective $T_m$.

Layering is in principle easy to understand, and simply represents the density response of
the liquid to vacuum. Like all response functions, it is intimately related
to the internal structure of the liquid, very thoroughly described in standard
textbooks\,\cite{tosi}. In short, if $g(r)$ is the liquid pair correlation
function, it will exhibit a main (damped) oscillation with some typical liquid
interparticle spacing $\lambda$. Its Fourier transform is the liquid static structure 
factor $S(k)$, which is correspondingly peaked around $ k_0 = 2\pi/\lambda$. 
When an external perturbation, say a $\delta$-function point-like repulsion, is inserted, the
liquid density will locally drop, and will recover away from the point with
a damped oscillation dictated by $S(k)$, and with a wavelength close to $\lambda$.
The surface represents another perturbation of that type (now extended) for the 
liquid metal, and layering is the result. Of course, the perturbation represented by 
vacuum is small only deep below the liquid surface, but is very large in the outermost layer.
Here the precise form of layering will generally take a different form from that suggested by
perturbative considerations. 

The metal atoms in the outermost surface layer are roughly speaking confined
in two dimensions. Moreover, in some cases---like in the heavy noble metals that
are inclined to surface reconstructions---the packing of surface atoms will
tend to be closer and tighter than the corresponding bulk atoms. The question therefore
arises whether the outer surface of a liquid metal could in some cases go as far
as crystallizing in two dimensions (2D). Such a crystallization would represent a case 
of ``surface freezing''---the opposite of surface melting---a phenomenon demonstrated 
and characterized in alkanes\,\cite{deutsch,deutsch1}. The additional interest of pursuing this
in elemental liquid metals would be the possibility to use this for a particularly
delicate test of the theories of two-dimensional freezing\,\,\cite{halperin-nelson}.

In two dimensions, freezing takes place in two steps. Disclination pairs
should bind into dislocations with a first transition, implying the onset of a ``hexatic phase''
with power law orientational order. Upon further cooling, a second transition binds 
dislocation pairs, with the onset of power law positional order. These
phenomena have been well documented in the melting of a 2D colloidal crystal\,\,\cite{murray}. 
A liquid metal surface might be able to exhibit, either above $T_m$
or in the supercooled state, a hexatic phase. Floating on top of 
a three dimensional liquid, the layered, nearly 2D solid surface would 
be able to exchange freely atoms with it, and disclinations/dislocations
could readily form and dissolve avoiding all the delicate kinetic 
problems presented by strict 2D systems.

Celestini {\em et al}\,\cite{celestini} carried out extensive MD simulations of the
liquid Au surface as a function of temperature particularly in the supercooling
regime. As Fig.\,\ref{celestcolor-fig}
shows, an increasingly strong layering is indeed expected,
and the top layer strongly resembles a 2D close packed lattice with a large number
of disclinations. Upon supercooling, disclinations rarefy and the system approaches
a hexatic transition. On the brink of that transition however, bulk crystallization 
suddenly takes place, starting precisely from the liquid surface inwards\,\cite{celestini1}.

An experimental attempt at pursuing experimentally the possible formation
of 2D hexatics at liquid metal surfaces would seem
very desirable, particularly with powerful tools such as surface X-ray reflectivity. 
Unfortunately, it appears that high melting point metals such as Au, Pt and Ir, where
layering and 2D crystallization tendencies are particularly strong, are technically
out of reach at least with this technique\,\cite{pershan}.

\section{Premelting of nanoclusters and nanowires}

Nanosystems became only recently a hot subject, but their thermal behavior 
has long been an important
one. Small clusters liquefy readily, or ``premelt'', well 
below the bulk melting temperature of the same material. Basic thermodynamics
predicts phenomenologically that a cluster of radius $R$ should melt at $T_m(R)$
\begin{equation} \label{eq8}
T_m(R) \thickapprox T_m \{ 1 - \frac{2}{\rho_l L R}\lbrack \gamma_{SV}-(\gamma_{SL}+\gamma_{LV})\rbrack\}
\end{equation}

This behavior was well verified experimentally in Au by classic cluster
beam experiments by Buffat and Borel\,\cite{borel}, whose results are shown
in Fig.\,\ref{borel-fig}.
A very parallel melting behavior is expected for one-dimensional extended nanosystems,
such as nanowires\,\cite{gulseren95}. In that case that Eq.~(\ref{eq8}) remains 
valid, although without the 
factor 2, as has also been confirmed by simulations. 

Microscopic theories as well as atomistic MD simulations of cluster 
melting are numerous and well documented. Reviews are given in standard 
books\,\cite{scolesbook,kumar}. For Pb clusters, the size-induced drop 
of $T_m(R)$ is clearly recovered by simulations (Fig.\,\ref{furiopoints}).
For Au clusters, of special experimental 
interest, MD simulations were carried out in order to understand how the premelted 
state was reached for increasing temperature. It was found first of all that
the experimental size-driven lowering of $T_m(R)$ for Au clusters is confirmed. 
It can be seen in Figs.\,\ref{furio-fig} and \ref{furio-fig1}, that 
liquid-like diffusivity sets in at the surface below the cluster melting temperature
(unless the cluster is
so small as to melt at $T_m(\infty)/2$ or below).
Melting at $T_m(R)$ occurs by a sudden propagation of this
liquid ``skin'' into the bulk-like interior.
This behavior is quite similar to that described by a phenomenological
model of Celestini {\em et al}\ \,\cite{celestini2}.

One would expect that clusters entirely bounded by  NM crystal faces 
should not premelt. While some overheating has been confirmed in Pb 
clusters bounded entirely by (111) facets\,\cite{metois}, this 
is not generally true. For example, clusters of systems with $H < 0$ 
such as In \cite{pavlovska} have been shown to premelt. The problem is
that Eq.\,(\ref{eq8}) is macroscopic, and as such it ignores microscopic details, such
as edges and corners that clusters nonetheless possess.

The simulated high temperature behavior of a NaCl cubic nanocluster ( a ``nanograin
of salt'') is quite revealing\,\cite{tanya3}. The (100) nanocube faces
are NM, and if they were unbounded, there would indeed be no premelting. However,
the cube corners are the weak spot, and begin 
to melt even below $T_m$. NaCl corner roughening was described long ago\,\cite{wortis}
however in a rigid step model, that did not permit melting.
It remains in principle an open question how the picture should evolve with 
increasing grain size. Clearly the corners will represent in all cases point-like
germs of the liquid which are present below $T_m$. While these germs will certainly
enlarge into liquid pools, it remains to be clarified how the pools will spread
out, and if and in what form their presence will permit, the 
faces of a large NaCl cube to exhibit NM and to remain solid up to and above $T_m$.
More generally, this approach should probably shed light also on the premelting
of nanoparticles made of materials with $H < 0$ mentioned above.
\section{Summary}
 We reviewed the high temperature behavior of solid surfaces, in particular their tendency 
 to wet spontaneously close to the bulk melting point. The case for which this does not happen 
 ( and the melt somewhat
 unexpectedly fails to wet its own solid---surface nonmelting) has been treated in considerable detail.
 The large variety of behavior shown by different classes of solids has been addressed. 
 We have also discussed other surface melting phenomena including rotational melting, glass surface defreezing,
 premelting of nanosystems. Finally, the existence of order at liquid surfaces has been highlighted as 
 an important theme that is attracting increasing interest. It is hoped that this panoramic overview, 
 despite its deliberate conciseness, will be of interest not only to the physics community, where it originates, 
 but also to the materials science and chemical physics community, where it might be of some use.  
\section*{Acknowledgments}\label{acknow}
This project was sponsored by Italian Ministry of University and Research, through
COFIN2003, COFIN2004, FIRB RBAU01LX5H, and FIRB RBAU017S8R; and by INFM, through PRA
NANORUB and ``Iniziativa Trasversale calcolo parallelo''.
A large fraction of the calculations reviewed here were performed at CINECA, 
Casalecchio (Bologna).



\clearpage

\clearpage
\clearpage
\clearpage
\newpage
\begin{figure}
\begin{center}
    \includegraphics[width=0.5\textwidth]{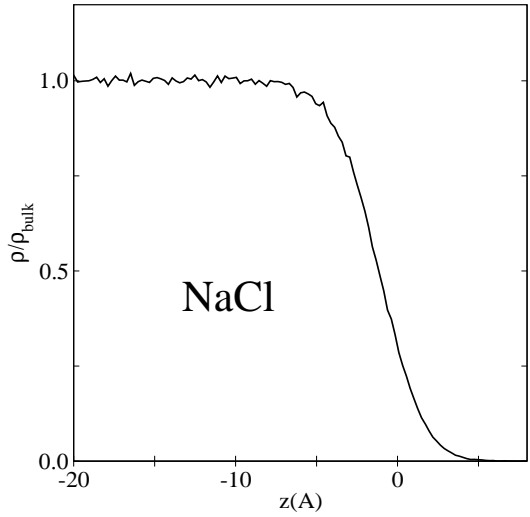}
    \caption[Density profile of the liquid surface of NaCl\ldots]{
            \label{NaClliq}
            Density profile of the liquid surface of NaCl obtained from a simulation at $T_m$. 
            The density drops very smoothly from the bulk in the liquid phase 
	    to that of the vapor one. 
	    From Ref.\,\protect\cite{tanya3}.}
\end{center}
\end{figure}

\begin{figure}
   \begin{center}
   \includegraphics[width=0.6 \textwidth, angle=-90]{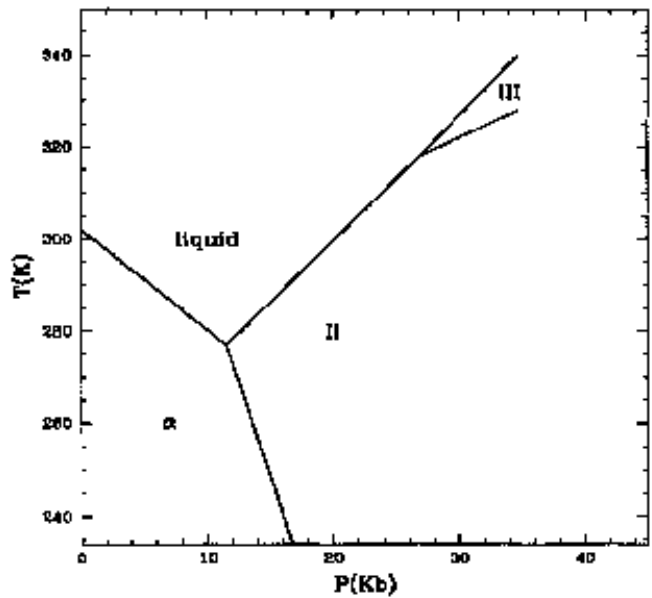}
   \caption[Schematic Ga phase diagram\ldots]{
           \label{ga-fig}
           Schematic Ga phase diagram, showing the semimetallic 
           $\alpha$-phase surrounded by fully metallic liquid and solid-II phases.
	   The picture for valence semiconductors such as Si, Ge is similar.}
 \end{center}
\end{figure}

\begin{figure}
\begin{center}
  \includegraphics[width=0.8 \textwidth]{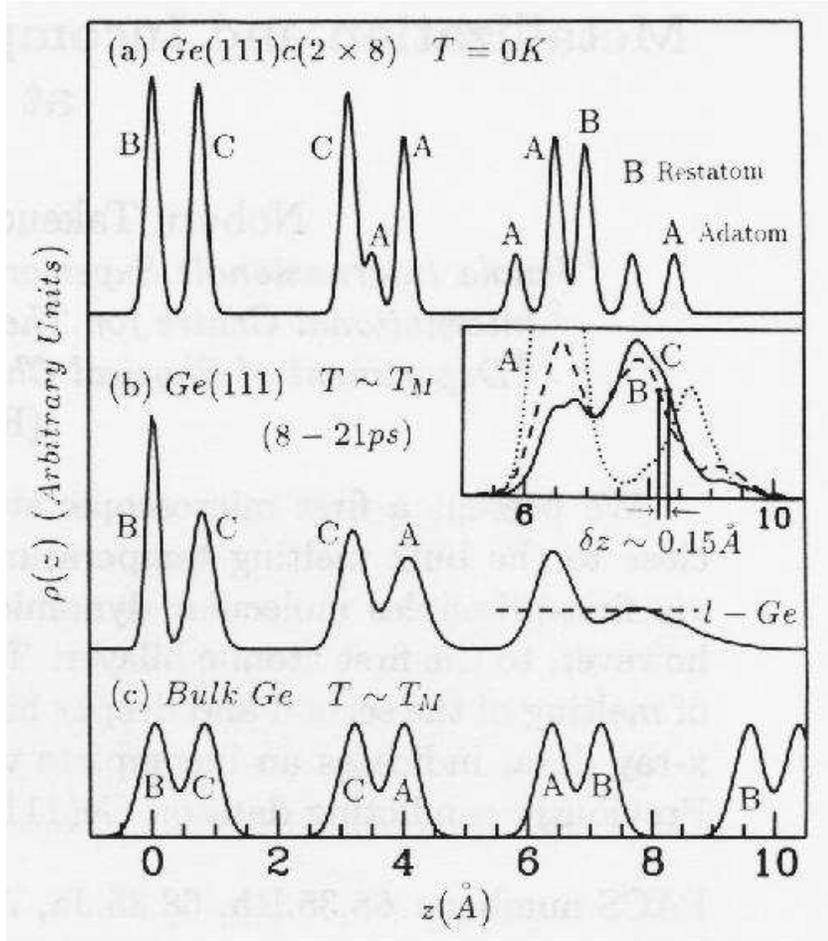}
  \caption[Averaged atom density profile for\ldots]{
            \label{takeuchi-fig}
            $(x,y)$ averaged atom density profile for (a) the Ge$(111)c(2 \times 8)$ reconstructed surface at $T = 0$K, 
            (b) the same Ge(111) surface at $T \sim T_M$, displaying ``blocked melting'' restricted to first bilayer 
	    and (c) crystalline bulk Ge at $ T \sim T_M$. Note in (b) 
	    the disordering of the first bilayer with atoms distributed between different sublattices (inset) 
	    From Ref.\,\protect\cite{takeuchi}.} 
\end{center}
\end{figure}

\clearpage
\begin{figure}
\begin{center}
   \includegraphics[width=\textwidth]{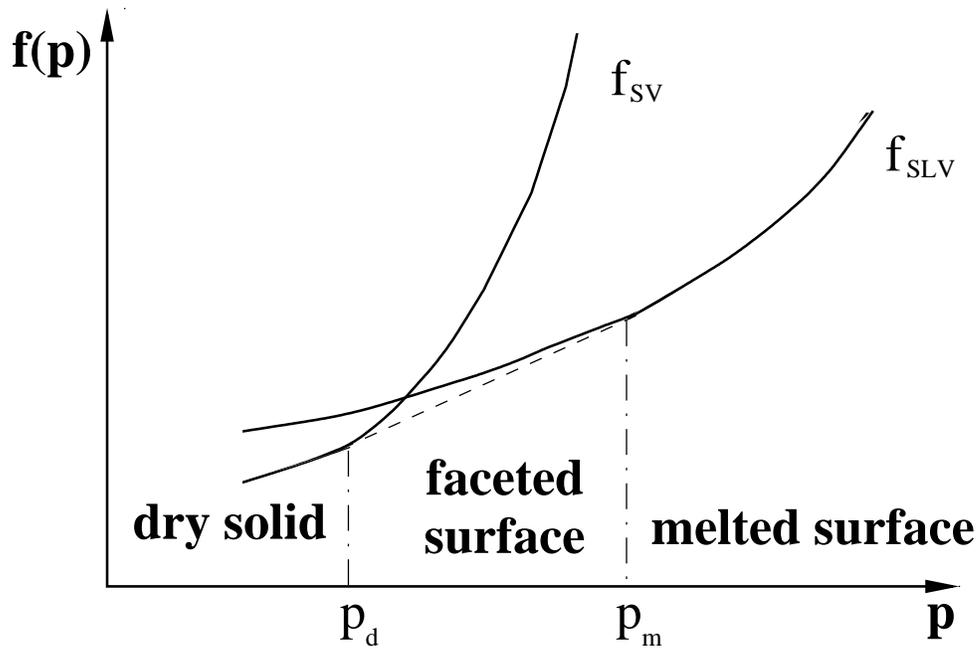}
    \caption[System with a solid branch and a liquid one\ldots]{
               \label{nozieres-fig}
               Plot of a system with a solid (non-melting) branch and a liquid one. 
               $p=\tan\theta$ and $f=\gamma/\cos\theta$, where $\theta$ is the tilting angle.
               The orientations up to $p_{d}$ remain \emph{dry}, in the range $(p_{d}, p_{m})$ there is 
               non-melting induced faceting, and above $p_{m}$ surface melting occurs. 
               $p_{d}$ might coincide with 0. From Ref.\,\protect\cite{nozieres}.}  
\end{center}
\end{figure}

\clearpage
\begin{figure}
 \begin{center}
   \includegraphics[width=\textwidth]{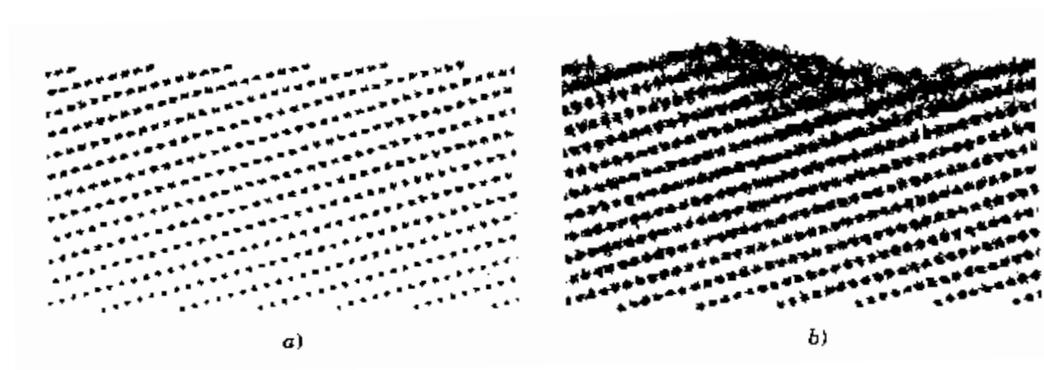}
   \caption[Non-melting induced faceting trajectories\ldots]{
            \label{bilal-fig}
            Non-melting induced faceting trajectories of the atoms in MD simulations, projected on a plane.
            (a) At low temperatures the Pb(432) surface is flat and stable. 
	    (b) At $0.97\, T_{m}$ the surface is phase separated in {\em dry} 
	    and {\em wet} 
	    facets. The atoms in the liquid part are recognizable by their 
	    wandering trajectories. From Ref.\,\protect\cite{bilalbegovic}.} 
 \end{center}
\end{figure}

\clearpage
\begin{figure}
 \begin{center}
    \includegraphics[width=0.8\textwidth]{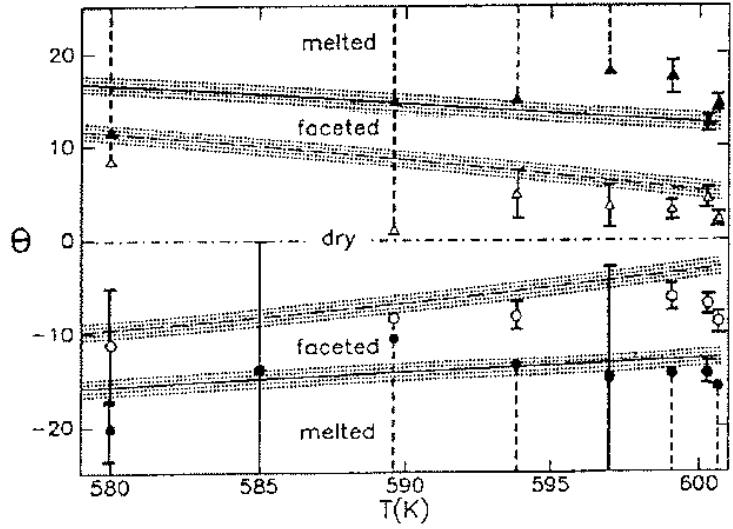}
    \caption[Non-melting induced faceting of Pb(111) vicinals\ldots]{
            \label{frenken-fig}
            Non-melting induced faceting of Pb(111) vicinals from Ref.\,\cite{frenkenfacet}. 
            Experimentally determined orientation angles as a function of $T$.
	    Triangles refer to a tilt toward the (110) orientation, while circles
	    refer to a  tilt toward the (100) orientation. 
	    Open symbols refer to the angle below which the vicinal is dry, 
	    while filled symbols to the angle above which the surface is melted. 
	    We refer to the original work 
	    for an explanation of the fit (solid and dashed curved).}
 \end{center}
\end{figure}

\begin{figure}
\begin{center}
   \includegraphics[width=0.8\textwidth]{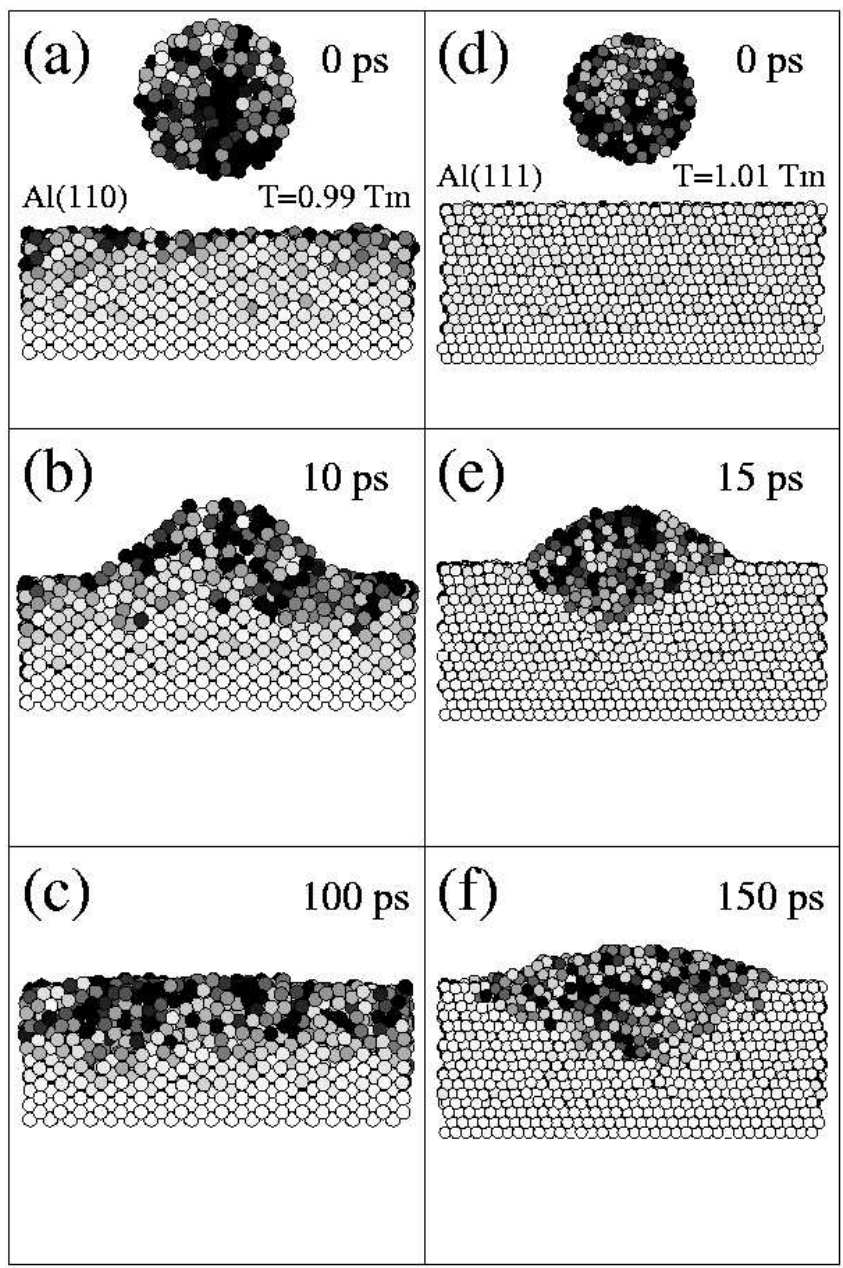}
   \caption[Evolution of Al liquid drop\ldots]{
            \label{ditolladrops-fig}
            Evolution of Al liquid drop on a surface of bulk Al. Left column:
            drop on a surface undergoing surface melting (Al(110) at 
	    $T = 0.99\, T_{m}$). (a) before contact;
            (b) after contact, the drop spreads readily;
	    (c) the drop has been fully absorbed. Right column:
            drop on a nonmelting overheated surface (Al(111) at $T= 1.01\,
	    T_{m}$). 
	    (d) before contact; 
            (e) after contact: the drop settles but does not spread; 
	    (f) final drop shape. Darkness of atoms proportional
            to their square displacement in the run.} 
\end{center}
\end{figure}

\begin{figure}
  \begin{center}
  \includegraphics[width=\textwidth]{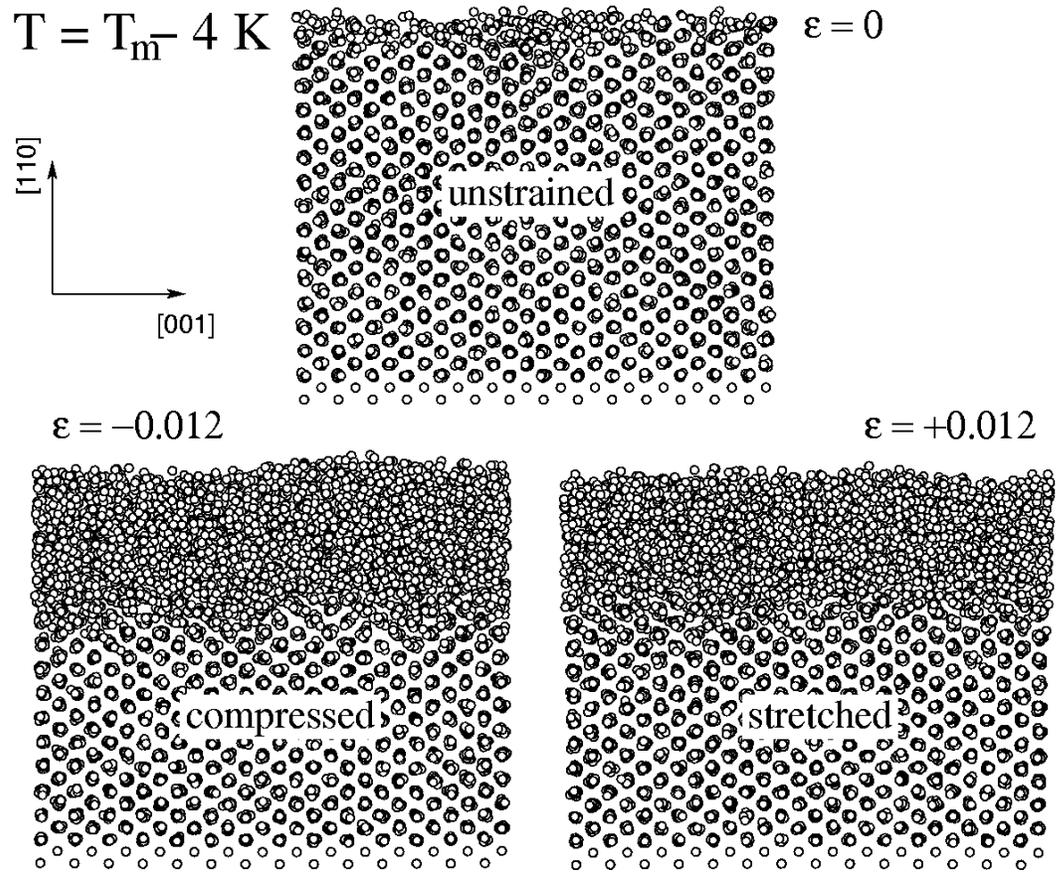}
  \caption[Simulated Al(110) surfaces under different strain\ldots]{
           \label{meltingstrain-fig}
           Single snapshots (lateral view) of three simulated Al(110) surfaces
           at the same temperature but under different strain conditions.
           Samples' size: $14\times 20\times 16$.
           The atom stacking in the 20-atom rows orthogonal to the picture
           clearly distinguishes the liquid and the solid phases.}
   \end{center}
\end{figure}
\clearpage

\begin{figure}
\begin{center}
  \includegraphics[width=\textwidth]{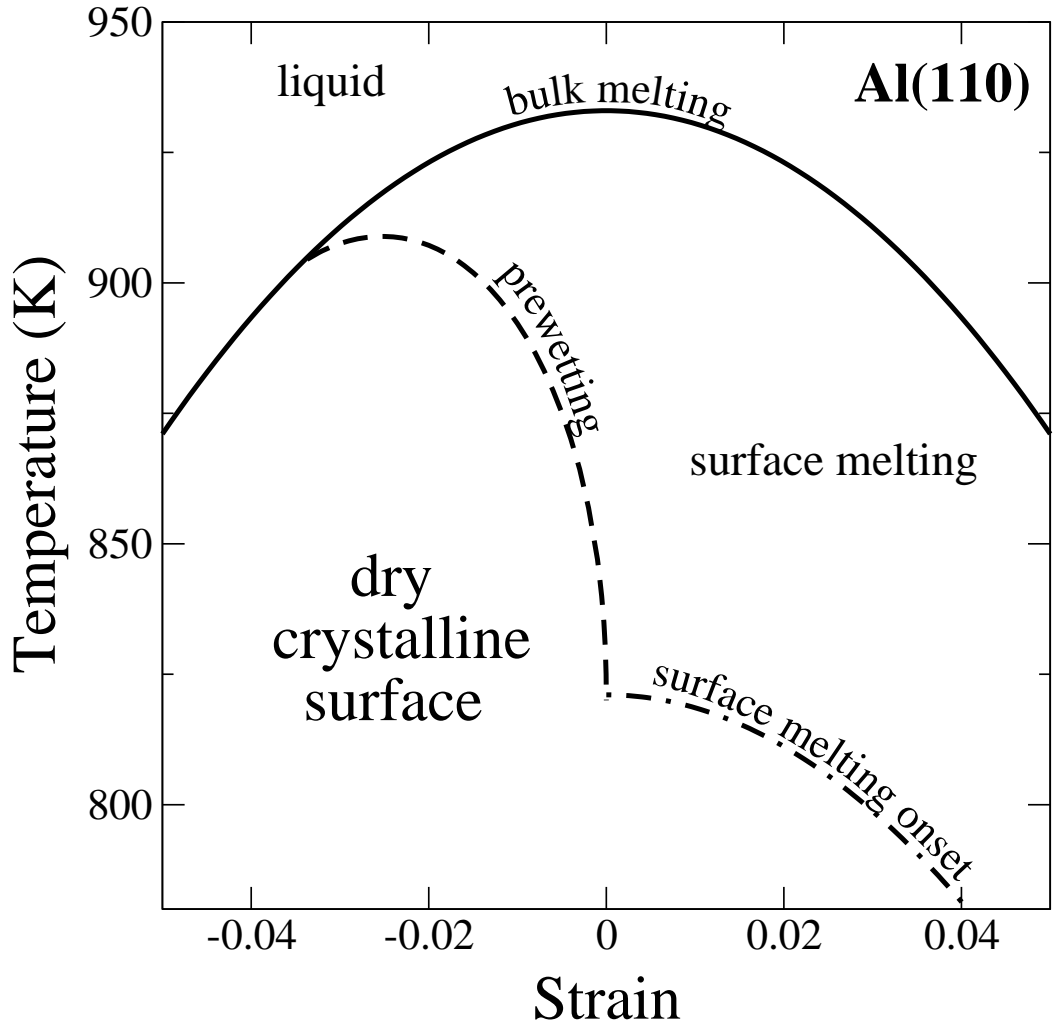}
  \caption[Phase diagram in presence of in-plane strain\ldots]{
  \label{diagram}
  Phase diagram of a metal surface in presence of in-plane strain. 
  Note that the negative (compressive) strain can give rise to a prewetting transition. 
  From Ref.\,\protect\cite{ugo}.} 
\end{center}
\end{figure}

\clearpage
\begin{figure}
 \includegraphics[width=1.1\textwidth]{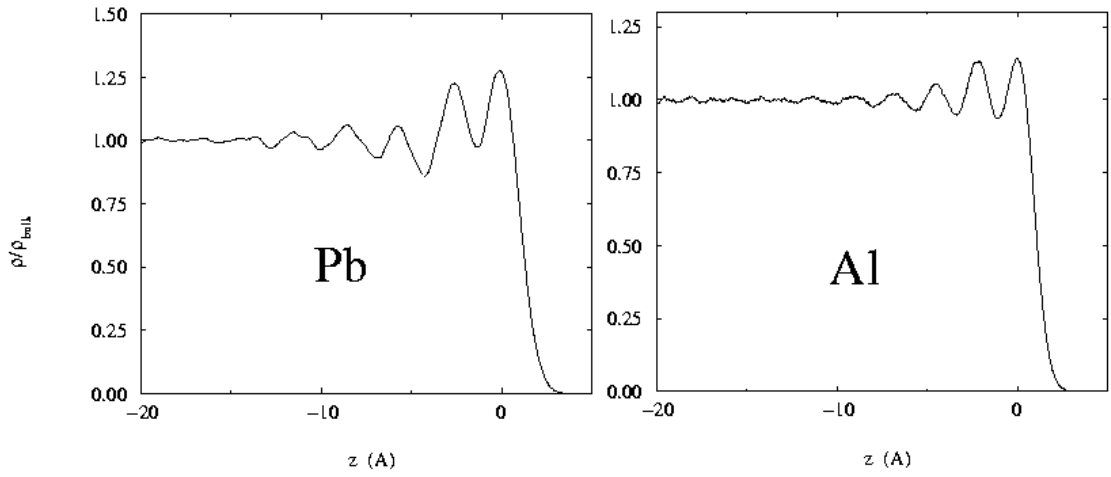}
 \caption[Density profile of the liquid surfaces of Pb and Al\ldots]{
            \label{pb-fig}
            Density profile of the liquid surface of Pb and Al obtained from a simulation at $T_{m}$.
            The profile shows layering oscillations at the liquid-vapor interface.} 
\end{figure}

 \begin{figure}
 \begin{center}
  \newpage
  \includegraphics[width=1.1\textwidth]{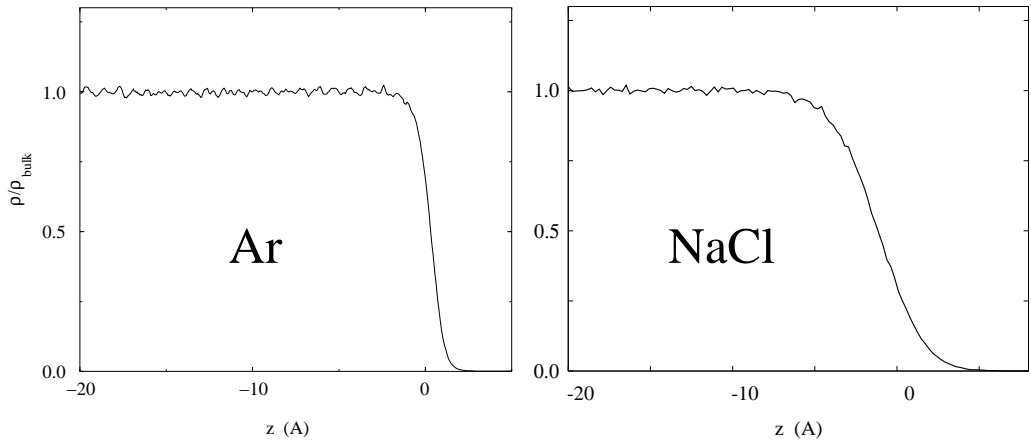}
   \caption[Density profile of the liquid surfaces of Ar and NaCl\ldots]{ \label{lj-fig}
            Averaged density profile of the liquid surface of LJ\,(Ar) and NaCl obtained from a simulation at $T_{m}$.
            The density drops smoothly from the bulk value in the liquid phase to that of the 
	    vapor phase, and the layering oscillations are undetectable.}
 \end{center}
 \end{figure}

\begin{figure}
  \clearpage
  \begin{center}
  \includegraphics[width=\textwidth]{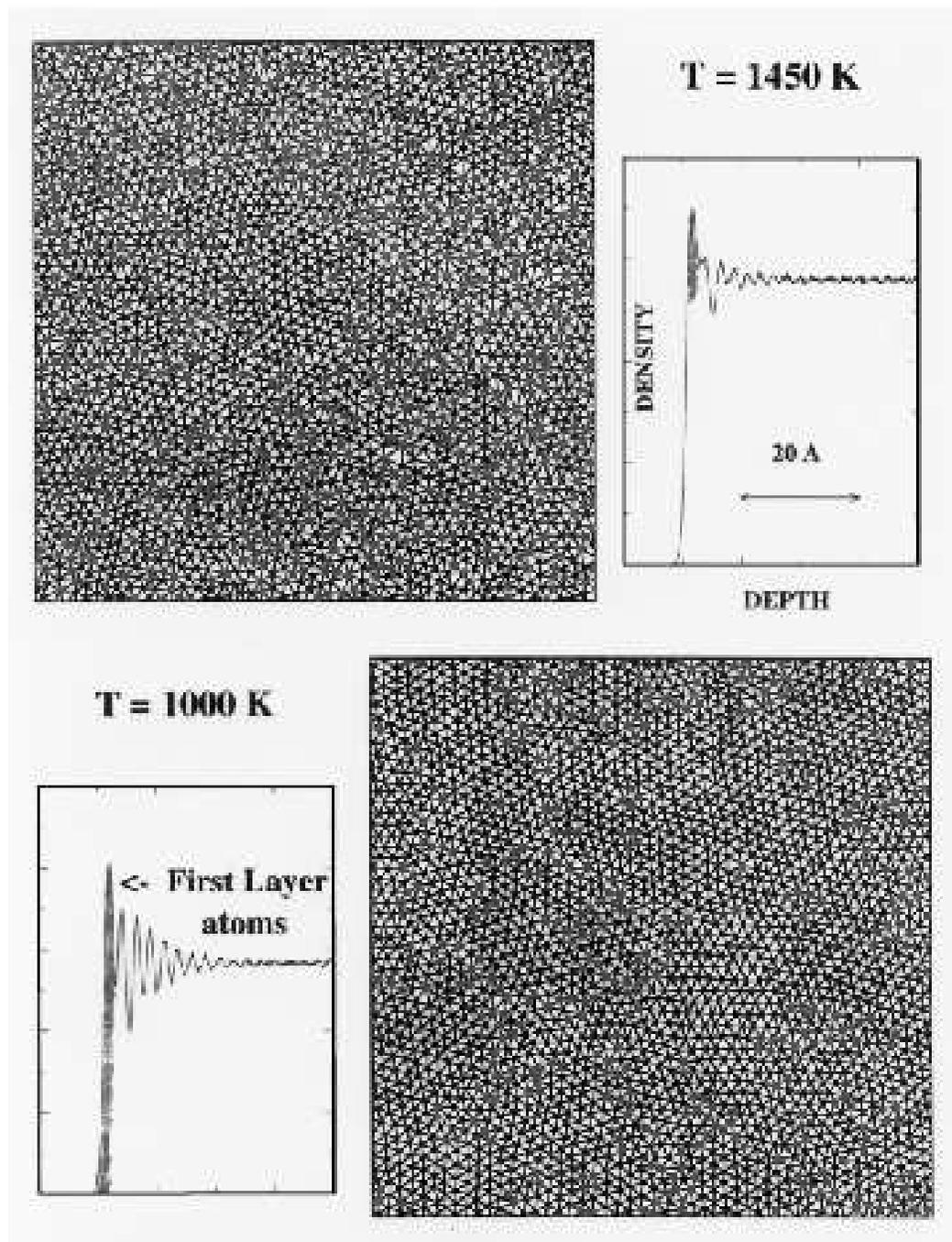}
  \caption[Nearly hexatic top layer of Au\ldots]{
            \label{celestcolor-fig}
            Nearly hexatic top layer of simulated liquid Au both above and 
            below (undercooling) the melting point $ T_m = 1335$K. 
	    Fivefold\,(green) and sevenfold\,(red) 
	    disclinations are pinpointed. The corresponding density profiles along the surface normal 
	    are also shown. The first peak corresponds to the outmost layer of atoms in the maps.
            From Ref.\,\protect\cite{celestini1}.}
\end{center}
\end{figure}

\clearpage
\begin{figure}
 \begin{center}
  \includegraphics[width=0.8\textwidth, angle = -90]{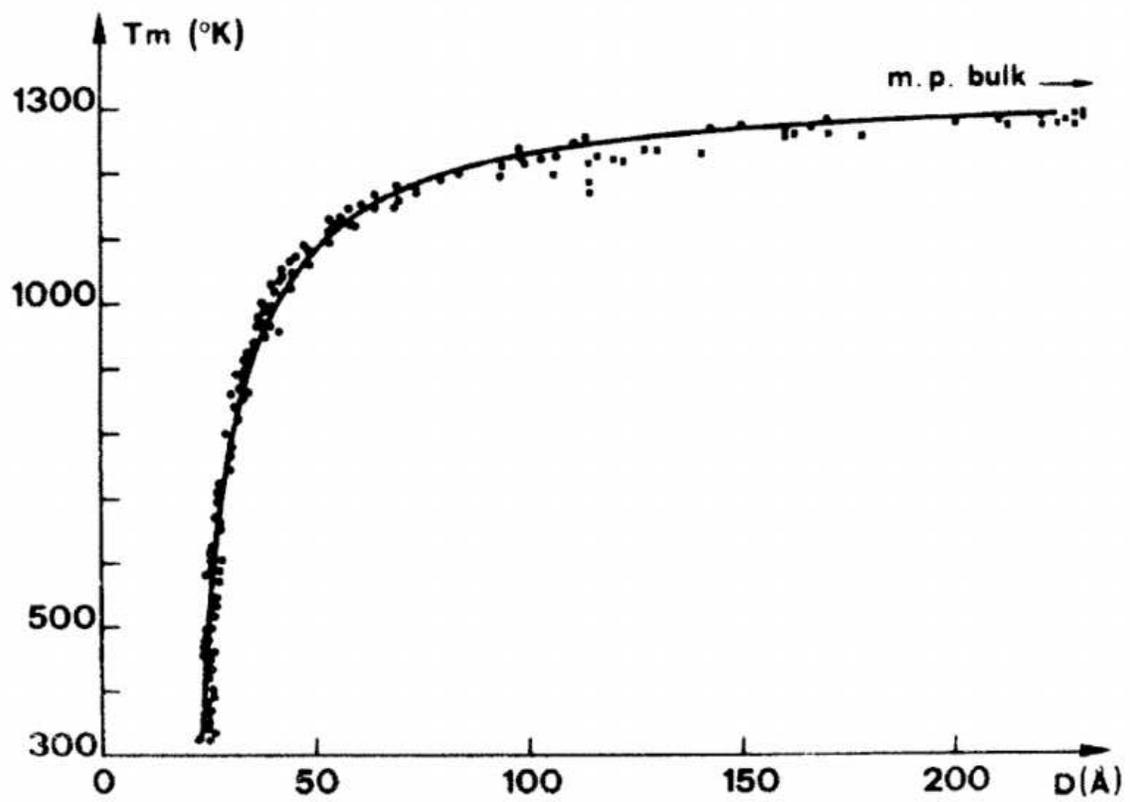}
    \caption[Melting point temperature of gold particles\ldots]{ \label{borel-fig}
            Experimental values of the melting point temperature of gold particles. 
            From Ref.\,\protect\cite{borel}.}
 \end{center}
\end{figure}

\clearpage
\begin{figure}
 \begin{center}
    \includegraphics[width=0.8\textwidth]{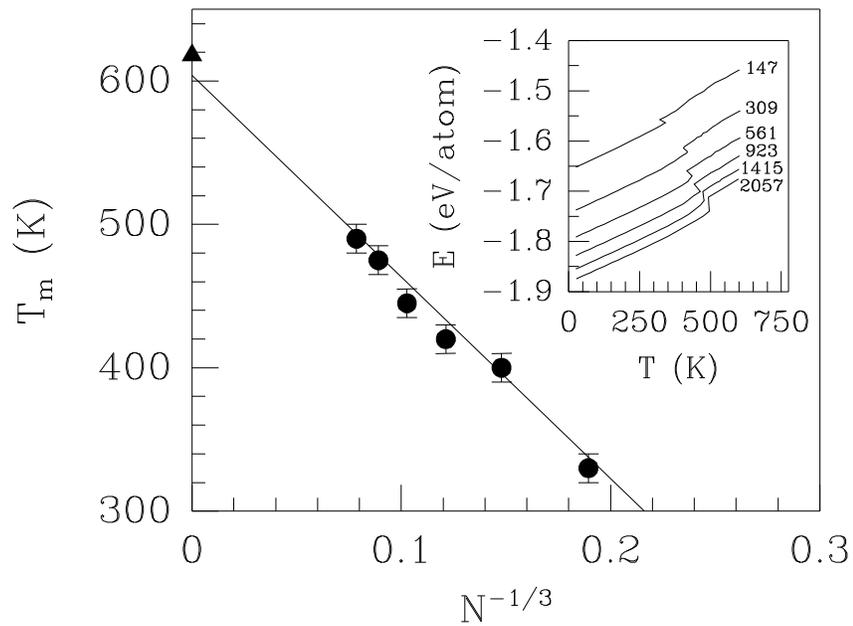}
    \caption[Melting temperature as a function of size\ldots]{
           \label{furiopoints}
           Melting temperature as a function of $N^{-1/3}$
           (where $N$ is the number of atoms) of simulated Pb
           clusters. The points correspond to the discontinuities in
           the caloric curves in the inset. The solid line is the linear
           fit to the points. The triangle corresponds to bulk melting.From Ref.\,\protect\cite{limong}}
 \end{center}
\end{figure}

\clearpage
\begin{figure}
 \begin{center}
  \includegraphics[width=0.6\textwidth, angle = -90]{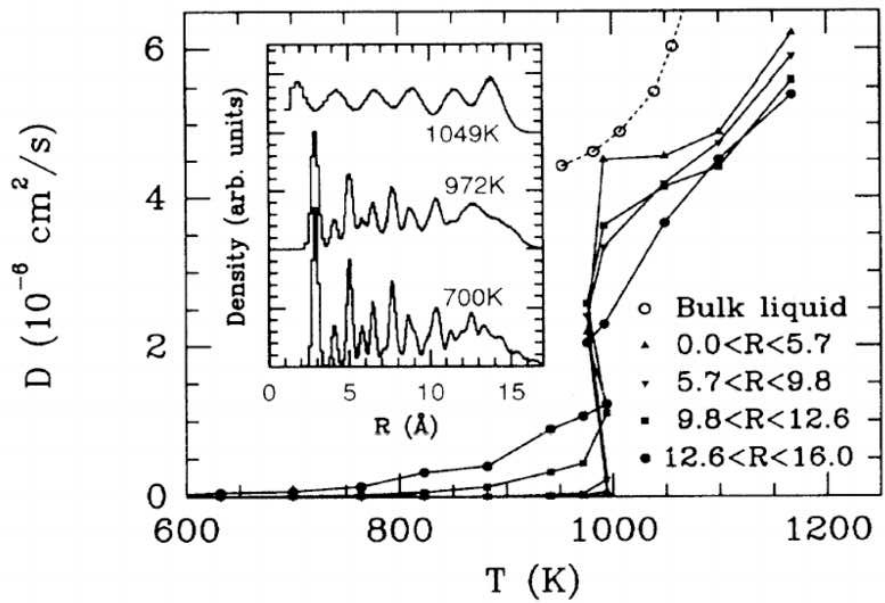}
  \caption[Au$_{879}$: $T$ dependence of the diffusion coefficient\ldots]{
            \label{furio-fig}
            Au$_{879}$: $T$ dependence of the diffusion coefficient for different shells, 
            compared with that for the supercooled bulk liquid. Inset:
	    Radial density distribution at T = 700, 972 and 1049K.}
 \end{center}
\end{figure}

\clearpage
\begin{figure}
 \begin{center}
 \includegraphics[width=0.55\textwidth, angle = -90]{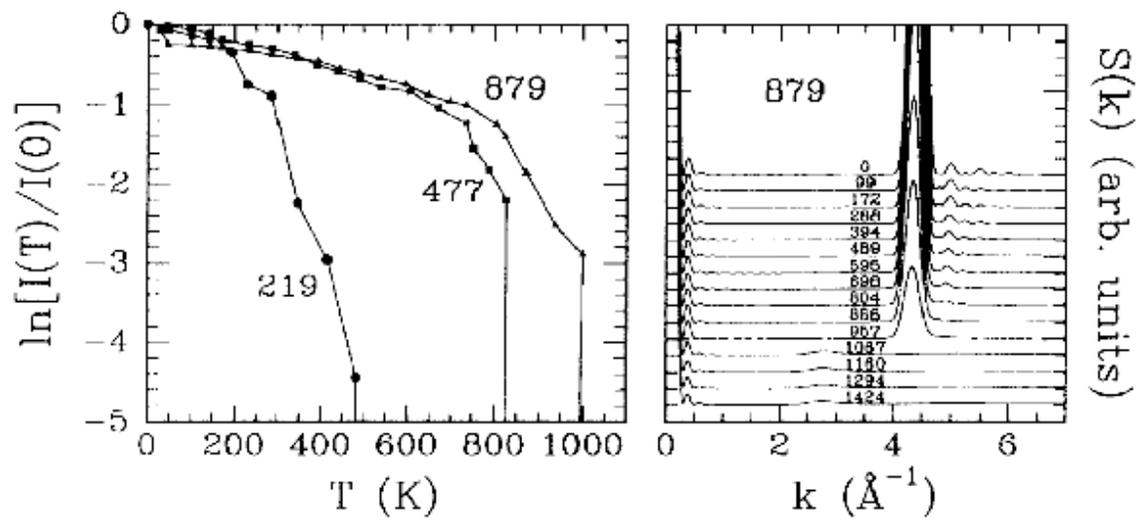}
 \caption[Logarithm of the effective Debye-Waller factor\ldots]{
            \label{furio-fig1}
            Logarithm of the effective Debye-Waller factor as a function of $T$
            for Au$_{219}$,  Au$_{477}$, and Au$_{879}$, and $T$ dependence of $S(k)$,
	    with $k$ along the (110) direction, for Au$_{879}$. From
	    Ref.\,\protect\cite{andreoni}. }
\end{center}
\end{figure}

\end{document}